\documentclass[fleqn,usenatbib]{mnras}

\usepackage{newtxtext,newtxmath}

\usepackage[T1]{fontenc}
\usepackage{ae,aecompl}
\usepackage{soul}

\usepackage{graphicx}	
\usepackage{amsmath}	
\usepackage{amssymb}	
\usepackage{bm}
\usepackage{xcolor}
\usepackage{multirow}
\usepackage{booktabs}
\usepackage{comment}
\usepackage[normalem]{ulem}
\usepackage{tabularx}

\newcolumntype{Y}{>{\centering\arraybackslash}X} 

\usepackage{amssymb}
\usepackage{pifont}

\definecolor{darkorchid}{rgb}{0.6, 0.2, 0.8}

\newcommand{\nodata}{$\cdots$}
\newcommand{\xmark}{\text{\ding{53}}}
\newcommand{\cmark}{\ding{51}}%
\newcommand{\jmr}{Journal of Marine Research}

\makeatletter\ifdefined\Hy@StartlinkName
\def\addOneNestingLevelStartLink{%
  \gdef\Hy@StartlinkName##1##2{%
    \sbox0{\Hy@StartlinkNameOrig{##1}{##2}}\usebox0
    \global\let\Hy@StartlinkName\Hy@StartlinkNameOrig%
  }%
}
\def\addOneNestingLevelEndLink{%
  \gdef\pdfendlink{%
    \sbox0{\pdfendlinkOrig}\usebox0%
    \global\let\pdfendlink\pdfendlinkOrig%
  }%
}
\let\Hy@StartlinkNameOrig\Hy@StartlinkName
\let\pdfendlinkOrig\pdfendlink
\else
\let\addOneNestingLevelStartLink\relax
\let\addOneNestingLevelEndLink\relax
\fi
\makeatother

\title[Vortex parameter study in self-gravitating discs]{
On the evolution of vortex in locally isothermal self-gravitating discs: a parameter study}

\author[D. Tarczay-Neh\'ez, K. Rozgonyi and Zs. Reg\'aly]{D. Tarczay-Neh\'ez$^{1,2}$\thanks{E-mail:tarczaynehez.dora@csfk.org}, K. Rozgonyi$^{1,3,4}$ and Zs. Reg\'aly$^{1}$\\
$^1$Konkoly Observatory, Research Centre for Astronomy and Earth Sciences, E\"otv\"os Lor\'and Research Network (ELKH), \\Konkoly Thege Mikl\'os \'ut 15-17, H-1121 Budapest, Hungary\\
$^2$MTA CSFK Lend\"ulet Near-Field Cosmology Group\\
$^3$International Centre for Radio Astronomy Research, The University of Western Australia, Crawley, WA 6009, Australia\\
$^4$Australian Research Council, Centre of Excellence for All-Sky Astrophysics in 3 Dimensions (ASTRO 3D), Australia
}

\date{Accepted XXX. Received YYY; in original form ZZZ}

\pubyear{2021}

\begin{document}
\label{firstpage}
\pagerange{\pageref{firstpage}--\pageref{lastpage}}
\maketitle

\begin{abstract}
Gas rich dusty circumstellar discs observed around young stellar objects are believed to be the birthplace of planets and planetary systems. Recent observations revealed that large-scale horseshoe-like brightness asymmetries are present in dozens of transitional protoplanetary discs. Theoretical studies suggest that these brightness asymmetries bf could be caused by large-scale anticyclonic vortices triggered by the Rossby Wave Instability (RWI), which can be excited at the edges of the accretionally inactive region, the dead zone edge. Since vortices may play a key role in planet formation, investigating the conditions of the onset of RWI and the long-term evolution of vortices is inevitable. The aim of our work was to explore the effect of disc geometry (the vertical thickness of the disc), viscosity, the width of the transition region at the dead zone edge, and the disc mass on the onset, lifetime, strength and evolution of vortices formed in the disc. We performed a parametric study assuming different properties for the disc and the viscosity transition by running 1980 2D hydrodynamic simulations in the locally isothermal assumption with disc self-gravity included. Our results revealed that long-lived, large-scale vortex formation favours a shallow surface density slope and low- or moderate disc masses with Toomre $Q \lesssim 1/h$, where $h$ is the geometric aspect ratio of the disc. In general, in low viscosity models, stronger vortices form. 
However, rapid vortex decay and re-formation is more widespread in these discs.
\end{abstract}

\begin{keywords}
accretion, accretion disc --- hydrodynamics --- instabilities --- methods: numerical --- protoplanetary discs 
\end{keywords}



\section{Introduction}
\label{sec:intro}

Since the first detection of an extrasolar planet around a main-sequence star by \cite{MayorandQueloz1995}, as of early 2021, more than 4300 exoplanets have been discovered. Three-quarter of the exoplanets are found in planetary systems (see, e.g. the Exoplanet Exploration Program and the Jet Propulsion Laboratory for NASA's Astrophysics Division\footnote{https://exoplanets.nasa.gov/}). The diversity of planetary systems urge the necessity of understanding the evolution of planets and planetary systems. Planet formation can be explained based on the core-accretion theory (see e.g. \citealp{Safronov1969}, \citealp{GoldreichandWard1973}, \citealp{Pollacketal1986}). In this scenario, planetesimals, the building blocks of planets, are build-up by the coagulation of dust particles. However, core-accretion theory suffers from the rapid loss of pebbles due to radial drift \citep{Weidenschilling1977}. An effective solution to this problem can be the development of dust traps formed at pressure maxima (see, e.g. \citealp{HaghighipourandAlan2003} and the references therein).

Recent theoretical studies revealed that anticyclonic vortices could form in protoplanetary discs. Vortices can be developed via the Rossby wave instability (RWI, \citealp{Rossby1939}. RWI is excited at a vortensity minimum of a steep pressure gradient in a protoplanetary disc \citep{Lovelaceetal1999}. Such places can occur at sharp viscosity transition regions, at which the magneto-rotational instability \cite[MRI, see,][]{BalbusandHawley1991} is switched off and on, \cite[see ][]{Lyraetal2015}, e.g., at the edges of a dead zone \citep{Gammie1996} or the walls of a gap opened by a giant planet \citep{devalBorroetal2007}. In the eye of an anticyclonic vortex, pressure maximum develops. Such places tend to trap dust particles
\citep[see more details in][]{BargeandSommeria1995,KlahrandHenning1997} which promotes the formation of planetesimals and planetary embryos \citep{Meheutetal2012}.

During disc life, the gas and dust material of protoplanetary discs are consumed. Transition discs are between the phase of a primordial gas- and dust-rich and gas- and dust-depleted phase. In this phase, the inner part of the disc material is depleted, while the outer parts of the disc contains a significant amount of gas and dust. Recent (sub)millimetre-wavelength observations revealed large structures like rings, inner cavities \cite[see e.g.][]{Andrewsetal2009} and horseshoe-like brightness asymmetries in about a dozen transition discs (see e.g. \citealp{Brownetal2009};
\citealp{vanderMareletal2013};
\citealp{Casassusetal2015};
\citealp{Marinoetal2015}; 
\citealp{Wrightetal2015};
\citealp{Andrewsetal2018};  \citealp{Maciasetal2018};    \citealp{Pinillaetal2019} and the references therein). Ring-like structures are also found by sub-millimetre observations done by ALMA of the DSHARP project \cite[see, e.g.,][]{Dullemondetal2018}. 

The origin of brightness asymmetries is still under a great debate. Although cavities in the gas are thought to be caused by an embedded massive planet \citep[see details in e.g.,][]{VanderMareletal2021} and photo-evaporation, while lopsided morphologies in the dust are thought to be caused by dust accumulation in the eye of large-scale anticyclonic vortices. \cite{Regalyetal2017} found intrinsic morphological differences between vortices formed at the outer edge of the dead zone and the edges of a wall of a cavity opened by a giant planet. They found that vortices formed at a giant planet opened gap wall are azimuthally less elongated and have higher azimuthal contrasts than they would have at the edge of a dead zone. Moreover, the vortex excited at the wall of a gap dissipates within $10^4 - 10^5$ years and lasts longer only in the case of a nearly inviscid disc. On the contrary, vortices formed at the dead zone edge can have an order of magnitude more extensive lifetime.

The gravitational stability of the disc can be described by the Toomre $Q$ parameter \citep{Toomre1964}. It was shown that the disc becomes gravitationally unstable to the axisymmetric perturbations when $Q\,\lesssim\,1$ in massive discs. Note that non-axisymmetric perturbations, i.e., spiral arms, can occur if the Toomre $Q$ parameter exceeds unity \citep[see, e.g.,][]{LauandBertin1978}. Hence, the mass of the disc is also a crucial point in modelling the dynamics of gas. Theoretical work of \cite{Lin2012} and \cite{LinandPapaloizou2011} revealed that sufficiently high disc mass could delay or even hinder large-scale vortex formation. \cite{Baeetal2015} found that large-scale vortices formed at the outer regions of protostellar discs tend to dissipate as $Q$ reaches unity. \cite{LovelaceandHohlfeld2013} and \cite{Yellinetal2016} found that self-gravity is important for discs with Toomre parameter $Q\,<\,Q_{\mathrm{crit}}\,=\,1/h$, where $h$ describes the geometric aspect ratio of the disc . As $Q$ reaches unity, large-scale vortices tend to dissipate.

Investigation of \cite{RegalyandVorobyov2017a} revealed that above a relatively low disc mass ($M_{\mathrm{d}}/M_{\star}~\geq~0.006$), self-gravity has an essential effect on the long-term evolution of large-scale vortices formed at sharp viscosity transitions. They showed that disc self-gravity stretches the vortex azimuthally, weakens and therefore shortens its lifetime, similarly to what was found by \citep{ZhuandBaruteau2016}.

Vortex enhanced planetesimal formation could be severely constrained for discs, in which no stable vortices can form, or large-scale vortex exists only on a comparatively short time scale. The lifetime of vortices might play a fundamental role in planet formation,  
as vortices are capable of collecting dust (the building blocks of planetesimals, see, e.g.,  \citealp{BargeandSommeria1995,KlahrandHenning1997,Meheutetal2012}). Hence, it is essential to bind the parameter space to determine the mass of stable vortex hosting discs, potentially helping planet formation. Moreover, constraining disc mass theoretically by simulations is also essential from the observational viewpoint.

In this paper, we investigate the effect of disc geometry (aspect ratio), the width of the viscosity transition region, disc mass and viscosity on the vortex formation, evolution and lifetime in self-gravitating and locally isothermal discs. We performed a parametric study, which includes 1980 models using a different wide range of disc parameters (see Table\,\ref{tab:inits}).

In Section\,\ref{sec:model}, we present our 2D hydrodynamic model. In Section\,\ref{sec:res}, we show our results on the role of initial density slope, disc geometry, and viscosity on the formation, evolution and lifetime of the vortices. Section\,\ref{sec:discussion} gives a discussion on our results and an estimation of the maximum allowable disc mass for vortex excitation. We conclude our results in Section\,\ref{sec:concl}. Appendix\,\ref{sec:app} deals with all model results, assuming a canonical reduction of viscosity inside the dead zone. Additional models assuming ten times stronger reduction in dead zone viscosity is available in the digital version. 

\begin{table*}
\caption{Parameters used for simulations. Here $\alpha_\mathrm{dz}$, $h$, and $p$ refer to the $\alpha$-parameter in the dead zone, the geometric aspect-ratio of the disc, and the surface density exponent, respectively. Columns $5$, $6$, and $7$ list $\Sigma_0$ values for a given $M_\mathrm{d}$ with $p=0.5,\,1.0$, and 1.5, respectively. The width of the transition  region, $\Delta r_\mathrm{dz}$, in nits of tyhe disc scale-height is given in column $8$. $\Delta r_\mathrm{dz}$ (in astronomical units) assuming $h=0.025,\,0.05$, and 0.1 are listed in columns $9$, $10$, and $11$, respectively. }

\label{tab:inits}
\begin{tabular}{lccccccccccc}
\hline\hline
\multirow{2}{*}{$\alpha_{\mathrm{dz}}$} & \multirow{2}{*}{$h$}  &   \multirow{2}{*}{$p$}  & $M_{\mathrm{d}}$ &   $\Sigma_{0, p=0.5}$  & $\Sigma_{0, p=1}$ & $\Sigma_{0, p=1.5}$ & $\Delta r_{\mathrm{dz}}$ &  $\Delta r_{\mathrm{dz,h=0.025}}$  &  $\Delta r_{\mathrm{dz,h=0.05}}$ &  $\Delta r_{\mathrm{dz
,h=0.1}}$ \\
&  &  & [$M_{\sun}$]  &  [$M_{\sun}/\mathrm{AU^2}$]  &  [$M_{\sun}/\mathrm{AU^2}$]  &  [$M_{\sun}/\mathrm{AU^2}$]  &   [H] &  [AU] & [AU] & [AU] \\
\hline\hline\noalign{\smallskip}

\noalign{\smallskip} 
$10^{-4}$   & 0.1   &  0.5  & 0.001   & $6.1144\cdot10^{-6}$ &   $3.1831\cdot10^{-6}$    &   $1.6111\cdot10^{-6}$   &   0.5  &   0.3 &   0.6 &   1.2   \\
$10^{-5}$   & 0.05  &   1   & 0.002   & $1.2229\cdot10^{-5}$ &   $6.3662\cdot10^{-6}$    &  $3.2221\cdot10^{-6}$    &   0.65  &   0.39    &   0.78    &   1.56   \\
            & 0.025 &   1.5 & 0.003   & $1.8343\cdot10^{-5}$ &   $9.5493\cdot10^{-6}$    &  $4.8332\cdot10^{-6}$    &   0.8  &   0.48    &   0.96    &   1.92   \\
            &       &       & 0.004   & $2.4458\cdot10^{-5}$ &   $1.2734\cdot10^{-5}$    &  $6.4443\cdot10^{-6}$    &   0.95 &   0.57    &   1.14    &   2.28   \\
            &       &       & 0.005   & $3.0572\cdot10^{-5}$ &   $1.5916\cdot10^{-5}$    &  $8.0554\cdot10^{-6}$    &   1.1 &   0.66    &   1.32    &   2.4     \\
            &       &       & 0.006   & $3.6686\cdot10^{-5}$ &   $1.9099\cdot10^{-5}$    &  $9.6663\cdot10^{-6}$    &   1.25  &   0.75    &   1.5     &   3      \\
            &       &       & 0.007   & $4.2801\cdot10^{-5}$ &   $2.2282\cdot10^{-5}$    &  $1.1278\cdot10^{-5}$    &   1.4 &   0.84    &   1.68    &   3.36   \\
            &       &       & 0.008   & $4.9815\cdot10^{-5}$ &   $2.5465\cdot10^{-5}$    &  $1.2889\cdot10^{-5}$    &   1.55  &   0.93    &   1.86    &   3.72    \\
            &       &       & 0.009   & $5.5029\cdot10^{-5}$ &   $2.8648\cdot10^{-5}$    &  $1.4500\cdot10^{-5}$    &   1.7 &   1.02    &   2.04    &   4.08  \\
            &       &       & 0.01    & $6.1144\cdot10^{-5}$ &   $3.1831\cdot10^{-5}$    &  $1.6111\cdot10^{-5}$    &   1.85 &   1.11    &   2.22    &   4.44   \\
            &       &       &         &                      &                           &                          &   2  &   1.2     &   2.4     &   4.8    \\

\hline\hline
\end{tabular}
\end{table*}

\section{Hydrodynamic model}
\label{sec:model}

We run 2D hydrodynamic simulations for the parameter study on the long-term vortex evolution in a protoplanetary disc. We used an improved version of the GPU supported the \textsc{gfargo}\footnote{http://fargo.in2p3.fr/-GFARGO-} code, which incorporates disc self-gravity. Hydrodynamical equations are solved on a 2D polar ($R,\phi$) grid in the locally isothermal approximation. 

The vertically integrated continuity and Navier-Stokes equations govern the dynamics and evolution of a locally isothermal protoplanetary disc, which read as follows

\begin{equation}
    \frac{\partial \Sigma}{\partial t}+\nabla \cdot (\Sigma\mathbf{v})=0,
\label{eq:cont}
\end{equation}
\begin{equation}
    \frac{\partial \mathbf{v}}{\partial t}+(\mathbf{v}\cdot \nabla)\mathbf{v}=-\frac{1}{\Sigma} \left( \nabla P+ \nabla{\cdot T} \right) -\nabla \Phi_\mathrm{tot},
\label{eq:NS}
\end{equation}

\noindent where $\Sigma$ and $P$ are the surface mass density and the vertically integrated pressure of the gas, respectively. $\mathbf{v}$ denotes the velocity of the gas, and $T$ is the viscous stress tensor whose components can are detailed in \cite{VorobyovandBasu2010}.

We include the disc self-gravity, hence the total gravitation potential ($\Phi_{\mathrm{tot}}$) contains the gravitational potential of the central star ($\Phi_{\star}$), and the disc itself ($\Phi_{\mathrm{sg}}$). As a large-scale vortex forms, the barycentre of the star-disc system is shifted from the centre of the grid, hence the indirect potential ($\Phi_{\mathrm{ind}}$) has also included in $\Phi_\mathrm{tot}$ (see its importance in, e.g., \citealp{MittalandChiang2015}, \citealp{ZhuandBaruteau2016} and \citealp{RegalyandVorobyov2017b}). Accordingly, 

\begin{equation}
\Phi_{\mathrm{tot}} = \Phi_{\star} + \Phi_{\mathrm{ind}} + \Phi_{\mathrm{sg}},
\end{equation}

where

\begin{equation}
\Phi_{\star} = -G\frac{M_{\star}}{r},
\end{equation}

\begin{equation}
\Phi_{\mathrm{ind}} = r\cdot G \int{\frac{\mathrm{d}m(\bm{r'})}{r^3}\bm{r'}},
\end{equation}
\begin{equation}
\label{eq:phisg}
\Phi_{\mathrm{sg}} = -G \int_{r_{\mathrm{in}}}^{r_{\mathrm{out}}}{r'dr} \times \int_0^{2\pi}{\frac{\Sigma\mathrm{d}\Phi'}{\sqrt{r'^2 + r^2 - 2rr' \cos{ (\Phi'-\Phi )}}}}.
\end{equation}

\noindent Here $r_{\mathrm{in}}$ and $r_{\mathrm{out}}$ are the inner and outer boundaries of the disc, while $\mathrm{d}m(\mathbf{r'})$ refers to the mass contained in a given grid cell. To solve equation~(\ref{eq:phisg}), Fast Fourier Transform technique \cite[see details in Section~2.8 in][]{BinneyandTremaine1987} was applied. To investigate disc fragmentation in gravitationally unstable protoplanetary discs, the same technique was successfully used in  \cite{VorobyovandBasu2010} and \cite{VorobyovandBasu2015}. \cite{RegalyandVorobyov2017a} also applied this technique for investigating vortex formation in self-gravitating discs.

As we use a locally isothermal assumption, the equation of state of the gas reads as
\begin{equation}
\label{eq:press}
    P = \Sigma c_{\mathrm s}^2,
\end{equation}
where $P$ and $c_{\mathrm{s}}$ are the local pressure and sound-speed of the gas, respectively. In a locally isothermal approximation, the local sound-speed can be given as 
\begin{equation}
\label{eq:chshomega}
c_{\mathrm{s}} = H\Omega,    
\end{equation}
where $H\,=\,hr$ is the local scale-height of the disc and $\Omega\,=\,\sqrt{GM_\star/r^3}$ is the Keplerian angular velocity. $G$ and $M_{\star}$ are the gravitational constant and the mass of the central star (both set to unity) at a given distance ($r$) measured from the central star.

To model the accretion of the gas (assumed to be driven by the magneto-rotational instability, MRI) onto the central star, for simplicity,we used the  $\alpha$-prescription of \cite{ShakuraandSunyaev1973}. In this assumption, the kinematic viscosity ($\nu$) of the gas is

\begin{equation}
\label{eq:shakura}
\nu\,=\,\alpha c_{\mathrm{s}}^2 / \Omega,    
\end{equation}
where $\alpha$ represents the effectiveness of MRI.

Formation of vortices can be excited at sharp viscosity transition developed in the edges of the dead zone. To describe the outer edge of a dead zone, we reduced the $\alpha$ parameter of the gas in the dead zone. As the transition region at the boundaries of the accretionally active and inactive zones is sharp \citep[see][]{Lyraetal2015}, this model is plausible to describe the edges of the dead zone. Note that utilising $\alpha$-prescription leads to a reduced kinematic viscosity in the dead zone also. For the reduction of $\alpha$ in the dead zone, ($\alpha_{\mathrm{dz}}\,=\,\delta_\alpha \alpha$), we used a steep $\alpha$ transition, which follows as

\begin{equation}
\label{eq:viscred}
\delta_{\alpha} = 1 - \frac{1}{2}\left ( 1 - \alpha_{\mathrm{mod}}  \right ) \left [1 - \tanh{\left ( \frac{r - r_{\mathrm{dz}}}{\Delta r_{\mathrm{dz}}} \right )} \right],
\end{equation}

\noindent where $\delta_\alpha$ reduces $\alpha$ by a factor of $\alpha_{\mathrm{mod}}$. The width of the transition region at the outer dead zone edge, $r_{\mathrm{dz}}$ is described by $\Delta r_{\mathrm{dz}}$. Note that, we modelled the outer edge of the dead zone only, as the inner edge lays well inside our computational domain. Note that with this model, the distance of the outer edge of the dead zone is fixed over time.

\subsection{Investigated disc models}

The inner and outer boundaries of the disc were set to $r_\mathrm{min} =3$ and $r_\mathrm{max}=50$ AU, respectively. The numerical resolution was $256$ in the radial and $512$ in the azimuthal direction. We used logarithmic in the radial and equidistant distribution of the grid cells in the azimuthal direction. In order to verify the numerical convergency, we run additional simulations with a numerical resolution of $512\times1024$ and $1024\times2048$. We found that our simulations with the standard configurations were numerically convergent.

We used wave damping boundary conditions of \cite{devalBorroetal2006} on both the outer and inner boundaries of the disc. Wave-killing zones are used at $r\leq1.2\cdot r_\mathrm{min}$ and $r\geq 0.9\cdot r_\mathrm{max}$ for each quantities over a time-scale of 10 orbits to the initial state. We note that with damping boundary conditions, the disc mass was not conserved within the simulation time. Instead, it increased by less than a percentage. All simulations were run for $10^5$ yrs, covering 1000 orbits at the distance of the vortex centre.

Initially, the surface density profile was set as a power-law function of distance

\begin{equation}
\label{eq:sig}
\Sigma = \Sigma_0 r^{-p},
\end{equation}

\noindent where $\Sigma_0$ is the surface density at 1 AU. Three different surface density exponent ($p$) were investigated ($0.5,\,1$ and $1.5$). We modelled 10 different disc masses between $0.001$ and $0.01\,M_{\sun}$ with a step of $0.001\,M_{\sun}$. The corresponding $\Sigma_0$ values for the given disc mass and $p$  can be found in Table\,\ref{tab:inits}. The effect of disc geometry was investigated by assuming three different values for $h$ (0.1, 0.05, 0.025). Note that the initial disk is not a quasy steady state solution in these models due to the initially applied viscosity reduction.

In order to investigate the effect of viscosity and the width of the viscosity reduction on large-scale vortex formation, we repeated all simulations with two different values of $\alpha_{\mathrm{dz}}$ ($10^{-4}$ and $10^{-5}$), while the global $\alpha$ parameter was set to $10^{-2}$ for both cases.

\begin{figure}
    \centering
    \includegraphics[width=\columnwidth]{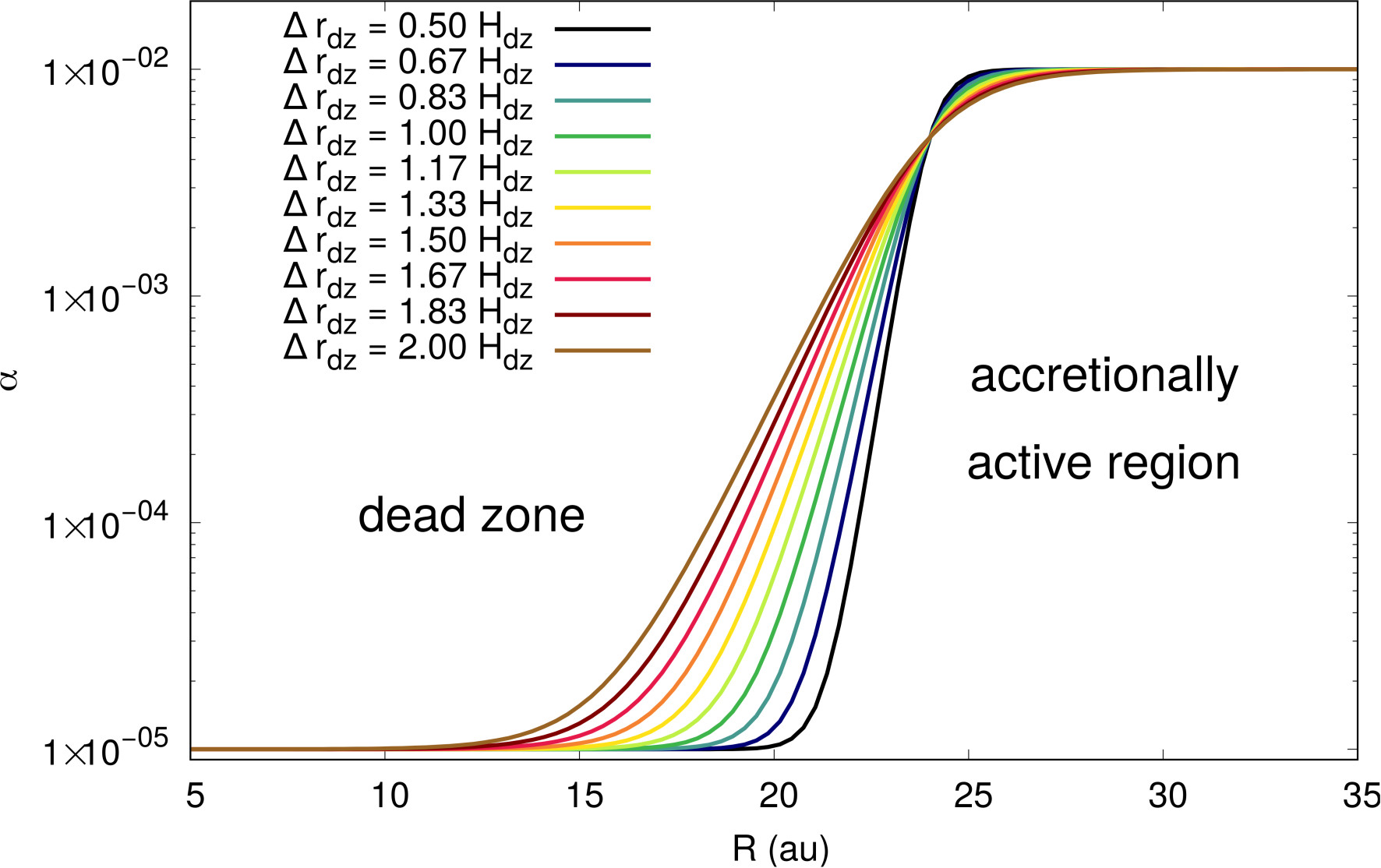}
    \caption{The width of the viscosity transition region at the outer edge of the dead zone at 24 AU in the case of $h\,=\,0.05$ and $\alpha_{\mathrm{dz}}\,=\,10^{-5}$. The ten different models assume  $0.5\,\mathrm{H_{dz}} \leq \Delta r_{\mathrm{dz}} \leq 2\,\mathrm{H_{dz}}$.}
    \label{fig:visc_trans}
\end{figure}

According to \cite{MatsumuraandPudritz2006}, the outer edge of the dead zone lies between 12 and 36 AU. Therefore, $r_{\mathrm{dz}}$ was set to $24$ AU in all simulations. The excitation of RWI requires sharp viscosity transition ($\Delta r_{\mathrm{dz}} \leq 2 H_{\mathrm{dz}}$, see \citealp{Lyraetal2009b} and \citealp{Regalyetal2012}). Therefore, we used 11 different values for $\Delta r_{\mathrm{dz}}$ between $0.5 - 2\,\mathrm{H_{dz}}$ with a resolution of $1/6\,\mathrm{H_{dz}}$ (see, e.g., Fig.\,\ref{fig:visc_trans} in which the viscosity transition is showed for $h\,=\,0.05$ models). Emphasize that $\alpha$ depends on $h$, therefore the steepness of viscosity transition differs in $h=0.025$ and $h=0.1$ models. Table\,\ref{tab:inits} shows the corresponding values of $\Delta r_{\mathrm{dz}}$ with the different aspect-ratios in astronomical units.

\begin{figure*}
    \centering
    \includegraphics[trim={0 50 0 2.7},clip,width=\textwidth]{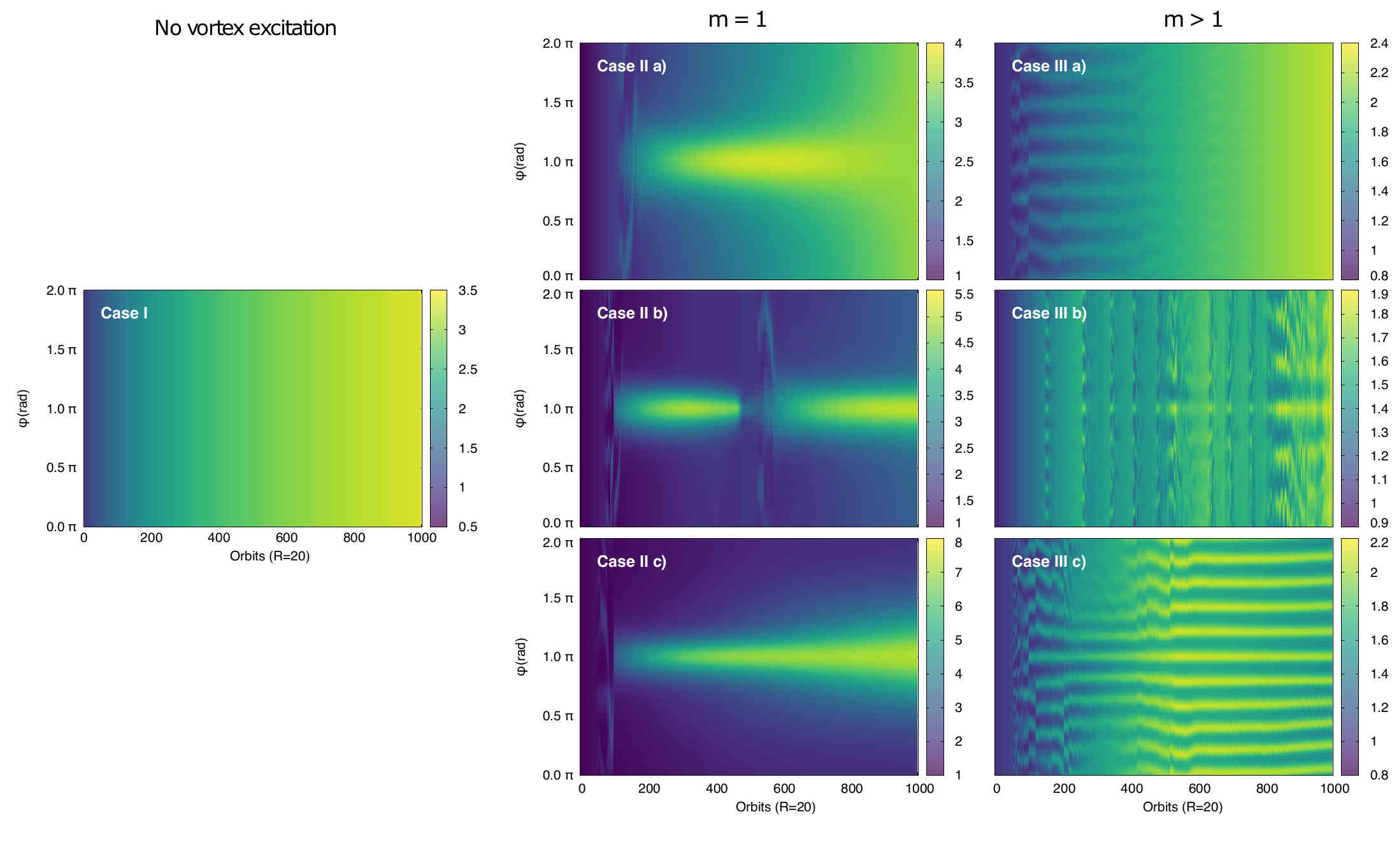}
    \caption{The evolution of $\delta \Sigma$ profile measured at the vicinity of the vortex eye in time in seven representative models. Time is measured in the number of orbits at the distance of the vortex eye. Case I represents those models in which RWI was not excited. Case II contains models where large-scale vortex formation ($m=1$) was occurred, while in Case III the coagulation of small-scale ($m>1$) vortices was inhibited. We further divided Class II and III into three subgroups: subgroup a) represents those models in which vortex formation and dissipation occurred (short-term vortices). Subgroup b) represents those models in which vortex formation, dissipation and re-formation were found. For Case II b) the vortex was sustained for about 200 orbits, then dissolved. After about 100 orbits, a new large-scale vortex was developed. Subgroup c) represents those models in which vortices lasted longer than our simulation time (long-term vortices).}
   \label{fig:trom}
\end{figure*}

\section{Results}
\label{sec:res}

As a pressure maximum develops at the outer edge of the dead zone, gas tends to accumulate there, forming a ring-like density enhancement. In general, RWI is excited, which results in vortex formation at the pressure maximum with a mode number of $m\,=\,3-6$. At later stages, small-scale vortices merge and form a single, large-scale anticyclonic vortex.

In general, the mode number is higher in low viscosity models (i.e., $\alpha_\mathrm{dz} = 10^{-5}$), than it is in high viscosity models ($\alpha_\mathrm{dz} = 10^{-4}$), independent of disc mass, which is consistent with what was found by \cite{RegalyandVorobyov2017a}. Moreover, $m$ depends on the $h$, i.e., the initial $m$ is the highest in $h=0.025$ models ($\sim5-6$), while it is $4-6$, and $3-4$ in $h=0.05$ and $h=0.1$, respectively, in the case of low viscosity models, independent of $p$, or $\Delta r_\mathrm{dz}$.

In order to investigate the effect of the kinematic viscosity, disc geometry and the width of the viscosity transition region on the long-term evolution of vortices, we calculated the vortex strength and the mean azimuthal density profile ($\delta \Sigma$) across the vortex eye. 

Vortex strength was measured as follows. First, the surface density distribution ($\Sigma$) was normalised by its initial distribution ($\Sigma^0$) at all time steps. In the next step, we fitted the 2D elliptical contours of the normalised surface density distribution ($\Sigma / \Sigma^0$) on a polar grid. For this, we assumed that the density distribution is elliptical inside the vortex, see, e.g., \citealp{Kida1981} and \citealp{Chavanis2000}). The aspect ratio of the fitted ellipse, $\chi_{\mathrm{dens}}$, was measured at the 87\% contour level of the maximum value of $\Sigma / \Sigma^0$. 

$\delta \Sigma$ was measured at the vicinity ($\pm 10$ grid cells in the radial direction) of the radial distance of the vortex eye. For this, we calculated the ratio of the maximum and minimum value of $\Sigma / \Sigma^0$. The evolution of vortices can be followed by repeating this procedure at each time step. For an example, the evolution of $\delta \Sigma$ in some representative models is shown in Fig.\,\ref{fig:trom}.

We found three distinct modes of evolution of the pressure jump. In Case I, no RWI excitation was observed. Case II represents models in which we observed large-scale vortex formation ($m=1$), while in Case III, only small-scale vortex formation was found ($m>1$), i.e., the vortex coagulation process was inhibited. 

We further divided Cases II and III into three subgroups. In Case II a) large-scale, while in Case III a) small-scale vortex formation occurred. For both cases, vortices lived only for a short-term. Cases II b) and III b) represent those models in which vortex re-formation (vortex formation after dissipation) occurred. In Case II c) and Case III c), the large- and small-scale vortices survived longer than our simulation time (i.e., long-term vortex formation was observed). Fig.\,\ref{fig:trom} shows an example for each Cases.

In the following three sections, we present our results for $\alpha_\mathrm{dz}=10^{-4}$ viscosity models. The effect of viscosity (by assuming, $\alpha_\mathrm{dz}=10^{-5}$) is presented in Section\,\ref{sec:low_visc}.

\begin{figure*}
    \centering
    \includegraphics[width=\textwidth]{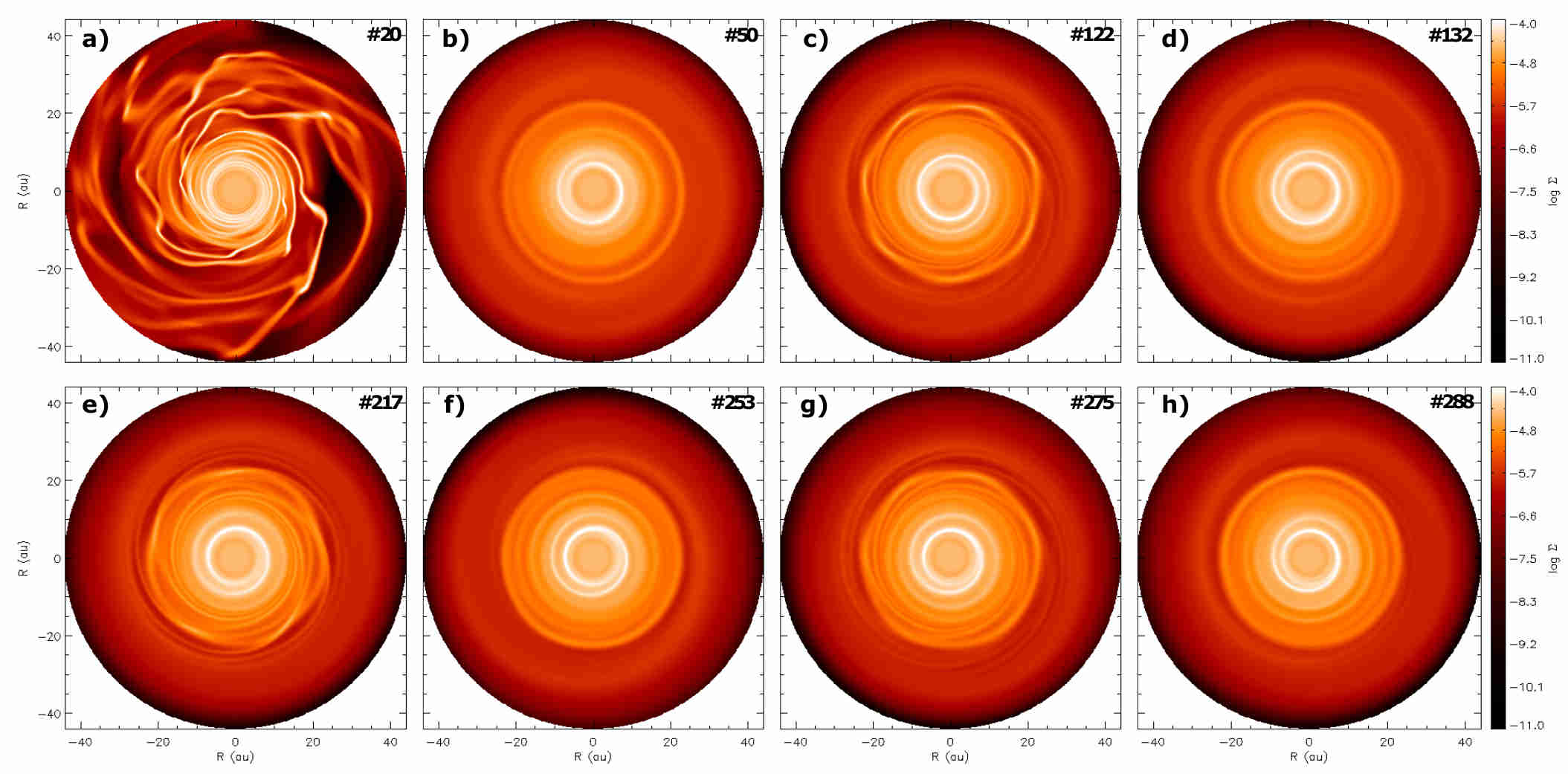}
    \caption{Evolution of the surface mass density distribution in $p=0.5$, $h=0.025$ high disc mass model ($M_{\mathrm{d}}/M_{\star} = 0.01$) assuming a sharp viscosity transition ($\Delta r_{\mathrm{dz}}=0.5H_\mathrm{dz}$).  The number of orbits is labelled in the right corner of each panel (e.g. panel a refers to the time-step at 20$^{th}$ orbits). After 20 orbits, the disc was fragmented into small scale clumps (see panel a). Clumps dissolved within 30 orbits and formed multiple ring-like structures (see panel b). After 70 orbits, the ring became RWI unstable, which resulted in the formation of multiple small-scale vortices (see panel c). Within less than ten orbits (see panel d), these vortices dissolved, and the ring-like structure reappeared. The process of vortex-ring transformation re-occurred at later stages. Emphasise that the coagulation of small-scale vortices was suppressed by the effect of self-gravity. Thus large-scale vortices could not be formed.}
    \label{fig:frag}
\end{figure*}

\subsection{Vortex evolution in \texorpdfstring{$p=0.5$}{p=0.5} models}

\subsubsection{\texorpdfstring{$h=0.025$}{h=0.025} simulations}
\label{sec:h0025_p05}

In the geometrically thin cases ($h=0.025$), large-scale vortex did not form. However, excitation of RWI could be observed below the medium mass models ($M_\mathrm{d} /M_{\star} \lesssim 0.005$), in which small-scale vortices formed, but their coagulation was inhibited. We also found that the wider the viscosity transition region was, the later the RWI excitation occurred (see Fig.\,\ref{fig:tromb_h0025_p05}). 

In low disc mass models that assumes steep viscosity transition region (e.g., $M_\mathrm{d} /M_{\star} = 0.001$, $\Delta r_{\mathrm{dz}} = 0.5 H_{\mathrm{dz}}$), long-lived but small-scale vortices formed (see panel of Case~III c) in Fig. \ref{fig:trom}). With increased disc mass, the lifetime of the vortices was shortened and vortex re-formation was observed (see panel of Case~III b) in Fig. \ref{fig:trom}).

The entire disc became gravitationally unstable and fragmented into small clumps within a few tenths of orbits ($\sim 10-20$, see panel a) in Fig.\,\ref{fig:frag}) for high disc masses ($M_\mathrm{d} /M_{\star} > 0.007$). However, at later stages, these clumps dissolved and formed a ring-like pressure bump at the dead zone edge (see panels b in Fig.\,\ref{fig:frag}). Although the disc was RWI unstable in this configuration, small-scale vortices could not be merged into a single large-scale vortex. The small-scale vortices could live for only a few orbits (about $10$ orbits for $M_{\mathrm{d}}/M_\star = 0.01$, $\Delta r_{\mathrm{dz}}=0.5H_{\mathrm{dz}}$, see, e.g., panels c in Fig.\,\ref{fig:frag}). The RWI excitation and vortex decay repeatedly occurred in this particular model. This indicates a series of formation and dissipation of multiple small-scale vortices with intermittent multiple ring structure. 

Here we note that ring-like structures seen on panels b, d, f, and h of Fig.\,\ref{fig:frag} are seemingly similar to the structures found by the DSHARP project analysing ALMA observations \cite[see, e.g.][]{Andrewsetal2018,Dullemondetal2018}. However, the ring-like structures seen in our simulations were developed in the gaseous component of the disc, while observations reflect the dust distribution. Therefore, further investigation taking into account the dust dynamics is required to explore this phenomenon in details. 

\subsubsection{\texorpdfstring{$h=0.05$}{h=0.05} simulations}

In low disc mass models ($M_\mathrm{d} /M_{\star} = 0.001$ and $0.002$), large scale vortex formed. The lifetime and strength of the large-scale vortex was dependent on disc mass and $\Delta r_{\mathrm{dz}}$. Namely, as the width of the transition region widened, the lifetime of the vortex shortened and also the $\delta \Sigma$ contrast weakened. 

In medium-disc mass models ($0.003 < M_\mathrm{d} /M_{\star} < 0.006$), small-scale vortices ($m>1$) formed. The lifetime of the vortices was shortened with increased disc mass and increased $\Delta r_{\mathrm{dz}}$. Note that re-formation of vortices was common in these models. The phase between two $m>1$ periods lasted longer at wider $\Delta r_{\mathrm{dz}}$ (see Fig.\,\ref{fig:tromb_h005_p05}). 

Large-scale vortices formed in models with $M_\mathrm{d} /M_{\star} \geq 0.006$ after 2-3 vortex cycles (see high disc mass panels in Fig.\,\ref{fig:tromb_h005_p05}). However, increased disc mass weakened the $\delta \Sigma$ contrast and also extremely shortened the lifetime of the large-scale vortex (vortex re-formation appeared within a few tenths of orbits). We also found that large-scale vortices in high disc mass models were elongated in the radial and azimuthal direction, leading to a highly eccentric disc. This phenomenon is caused by the indirect term. See details in Section\ref{sec:vort_split} in the Discussion.

\subsubsection{\texorpdfstring{$h=0.1$}{h=0.1} simulations}

Although RWI was excited in these models assuming low disc masses, large-scale vortex formation was inhibited due to the stretching effect of self-gravity. To resolve the apparent contradiction, see details in Section\,\ref{sec:vort_split} in the Discussion.
Large-scale vortices were stretched within a short period (a few tenths of orbits), and a ring-like structure formed. In high disc mass models, the mass of the accumulated gas in the vortex has grown sufficiently large to shift the barycentre of the star-disc system. Therefore, the disc tended to wobble around the barycentre. This led to a highly elongated vortex shape in the radial direction.

\subsection{Vortex evolution in \texorpdfstring{$p=1$}{p=1} models}

\subsubsection{\texorpdfstring{$h=0.025$}{h=0.025} simulations}

In low mass models ($M_\mathrm{d}/M_{\star} \lesssim 0.003-0.004$), large-scale vortex formation occurred, assuming $\Delta r_{\mathrm{dz}} < 1.7H_\mathrm{dz}$. The coagulation process of small-scale vortices did not occur within the simulation time in the middle- ($0.003 - 0.004 \lesssim M_\mathrm{d}/M_\star \lesssim 0.006-0.007$) and high-disc mass cases ($M_\mathrm{d}/M_\star \geq 0.007$). Moreover, increased disc mass narrowed the $\Delta r_{\mathrm{dz}}$ range in which RWI excitation could occur. 

In high disc mass models, vortices formed only in sharp viscosity transition models (e.g. assuming $M_{\mathrm{d}}/M_\star = 0.01$, RWI excitation occurred if $\Delta r_{\mathrm{dz}} < 0.95H_{\mathrm{dz}}$.) 

\subsubsection{\texorpdfstring{$h=0.05$}{h=0.05} simulations}

In $p=1$ models, small-scale vortices were able to coagulate. Hence large-scale vortex developed in these models. The strength and lifetime of the large-scale vortex depend on $M_{\mathrm{d}}$ and $\Delta r_{\mathrm{dz}}$. Namely, increased disc mass or width of the viscosity transition region weakened the vortex and shortened its lifetime.

\subsubsection{\texorpdfstring{$h=0.1$}{h=0.1} simulations}
\label{sec:3.3.3.}

Large-scale vortex formation occurred in $h=0.1$ cases. However, we found that RWI excitation is $\Delta r_{\mathrm{dz}}$ limited within the investigated  $\Delta r_{\mathrm{dz}}$ - $M_{\mathrm{d}}$ range.  Independent of disc mass, discs beyond $\Delta r_{\mathrm{dze}} > 1.4H_{\mathrm{dz}}$ were RWI stable (see Fig.\,\ref{fig:tromb_h01_p1}).

Moreover, in low-disc mass models, vortex re-formation occurred. 
We also found that the contrast $\delta \Sigma$ (and hence the vortex strength) is directly proportional to the disc mass: the higher the disc mass, the stronger the contrast.

\begin{figure*}
    \centering
    \includegraphics[trim={0 190 0 0},clip,width=0.95\textwidth]{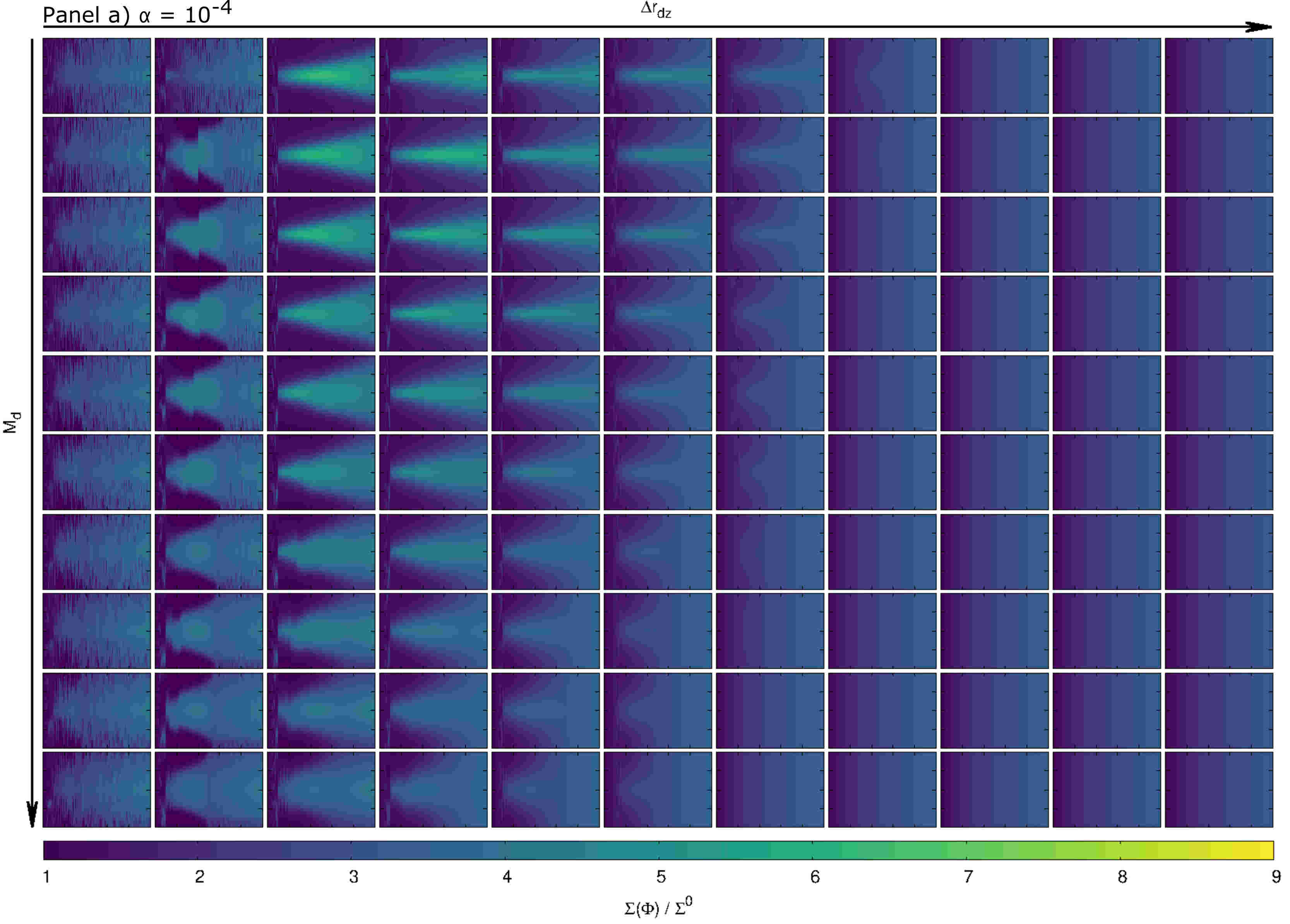}\vspace{2em}
    \includegraphics[width=0.95\textwidth]{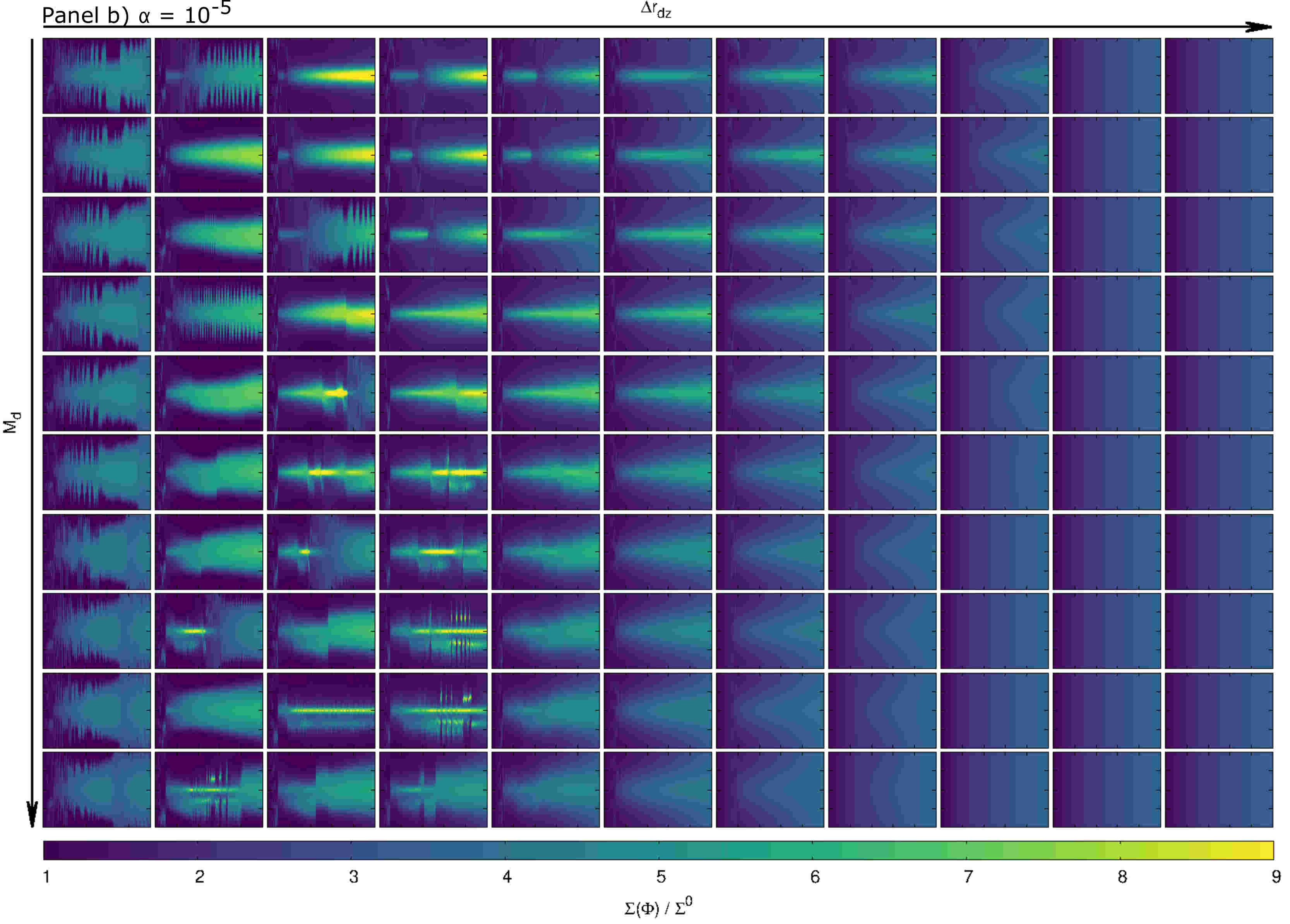}
    \caption{Evolution of $\delta \Sigma$ in $h = 0.05$, $p=1$ models. Panel a refers to $\alpha = 10^{-4}$ cases, while panel b refers to $\alpha = 10^{-5}$ models. The width of the transition region widens, and the disc mass increases from left to right and from top to bottom, respectively.}
    \label{fig:tromb_h005_p1}
\end{figure*}

\begin{figure*}
    \centering    \includegraphics[width=\textwidth]{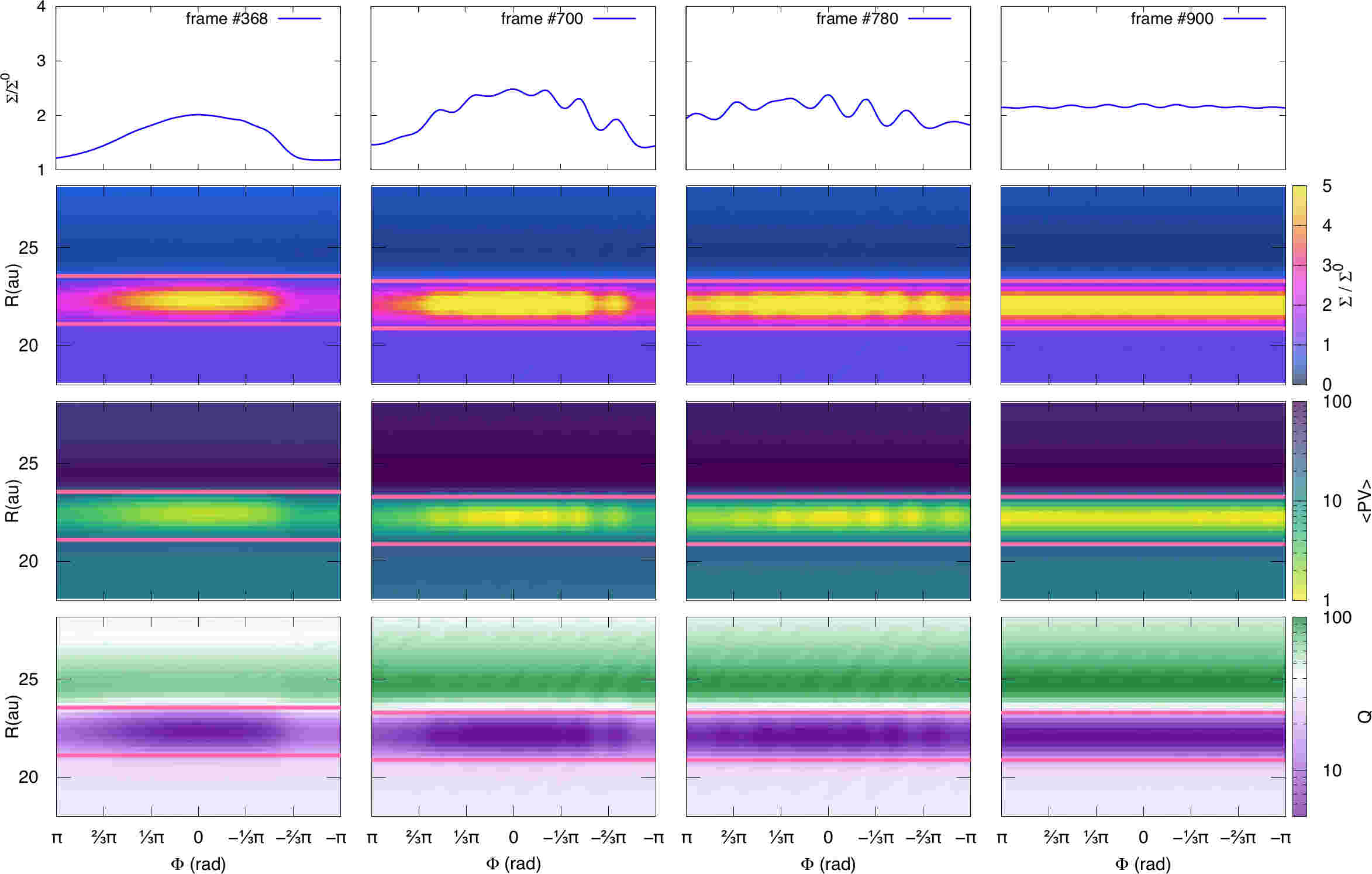}
    \caption{Vortex splitting in a particular model in which $h=0.025$, $p=1$, $\Delta r_{\mathrm{dz}} = 0.95H_\mathrm{dz}$ and $M_{\mathrm{d}}/M_{\star} = 0.004$ were used. The different columns refer to the different stages of the azimuthal profile of $\delta \Sigma$, PV and the Toomre $Q$. From left to right, frames $\#368$, $\#700$, $\#780$ and $\#900$ are shown. Pink horizontal lines on the panels refer to the region in which the radial average of $\Sigma / \Sigma^0$ was calculated.}
    \label{fig:vort_split}
\end{figure*}

\subsection{Vortex evolution in \texorpdfstring{$p=1.5$}{p=1.5} models}

\subsubsection{\texorpdfstring{$h=0.025$}{h=0.025} simulations}

Contrary to $p=0.5$ and $1$ cases, large-scale vortices developed in those $p=1.5$ models where RWI was excited, independent of the disc mass.
Vortices formed in $p=1.5$ models were stronger (the $\delta \Sigma$ contrast was higher, thus less elongated in the azimuthal direction) than in  $p=1$ models. For example, $\delta \Sigma \simeq 6.5$ for $p=1.5$ models, while $\delta \Sigma \simeq 4$ for $p=1$ models in low disc mass cases, see Figs.\,\ref{fig:tromb_h0025_p1} and \ref{fig:tromb_h0025_p15}. We also found that increasing disc mass or $\Delta r_{\mathrm{dz}}$ weakened and shortened the lifetime of the large-scale vortex. Moreover, RWI excitation was $\Delta r_{\mathrm{dz}}$ limited, similar to $p=1$, $h=0.1$ models. Assuming $\Delta r_{\mathrm{dz}} > 1.4H_{\mathrm{dz}}$, the disc was RWI stable.

\subsubsection{\texorpdfstring{$h=0.05$}{h=0.05} simulations}

In these models, large-scale vortex developed in all cases in which RWI was excited. However, RWI excitation was $\Delta r_{\mathrm{dz}}$ limited, similar to $h=0.025$ models. Discs having $\Delta r_{\mathrm{dz}}$ above $1.4H_{\mathrm{dz}}$ were RWI stable (see Fig.\,\ref{fig:tromb_h005_p15}), hence vortex formation was absent. 

\subsubsection{\texorpdfstring{$h=0.1$}{h=0.1} simulations}

Similar to $p=1$ models, RWI excitation was $\Delta r_{\mathrm{dz}}$ limited in $p=1.5$ models: we did not observe RWI excitation in simulations where $\Delta r_{\mathrm{dz}} > 1.25H_{\mathrm{dz}}$ (see Fig.\,\ref{fig:tromb_h01_p15}).

\subsection{The effect of viscosity}
\label{sec:low_visc}

As a general effect of the viscosity, vortex formation could be triggered at wider transition region of viscosity in low viscosity ($\alpha_\mathrm{dz}=10^{-5}$) models, see, e.g., upper ($\alpha=10^{-4}$) and lower ($\alpha=10^{-5}$) panels of Fig. \ref{fig:tromb_h005_p1}. $\Delta r_\mathrm{dz, crit}$ is the critical value for $\Delta r_\mathrm{dz}$, below which RWI excitation occurred, which is a necessary condition to form large-scale vortices. Assuming $p=1$, $h=0.05$, and the lowest disc mass in $\alpha_\mathrm{dz}=10^{-4}$ and $10^{-5}$ models, $\Delta r_\mathrm{dz, crit}$ were $1.4$ and $1.7H_\mathrm{dz}$, respectively (see panels a and b in Fig.\ref{fig:tromb_h005_p1}). This means that the maximum value for $\Delta r_\mathrm{dz}$ at which RWI can be excited was $0.3H_\mathrm{dz}$ wider in low viscosity models. This trend holds for $h=0.025$, and $0.05$ models too, however, the difference was $0.15H_\mathrm{dz}$ in $h=0.1$ models.

It can also be seen in the panels a and b of Fig. \ref{fig:tromb_h005_p1} that viscosity affects the contrast in $\delta \Sigma$. In low viscosity models, azimuthally less elongated, i.e., stronger vortices formed. Viscosity also affects the vortex lifetime: in general, vortices were sustained longer in low viscosity models. However, vortex re-formation occurred mainly in low viscosity models (see e.g. low disc mass models on the lower panels of Fig.\,\ref{fig:tromb_h005_p1}).

Regarding the jumps appreciable in the second columns on both panels of Fig.\,\ref{fig:tromb_h005_p1}) we found that they were caused by the algorithm applied to determine the centre of the vortex. Namely, the local maximum of $\delta \Sigma$ can be moved sometimes quite fast.

For a more comprehensive view of the effect of viscosity, additional figures of low viscosity models are available in the online material.

\section{Discussion}
\label{sec:discussion}

\subsection{Vortex splitting and disc stability}
\label{sec:vort_split}

After fully developed, large-scale vortices tend to split up into small-scale vortices (two, three or more), mostly in low disc mass models assuming a disc aspect ratio of $h=0.05$ and $0.025$ and a surface density slope $p=0.5$, $1$ (see, e.g., models assuming $M_\mathrm{d}/M_\star \lesssim 0.004M_\odot$ and $\Delta r_\mathrm{dz}\lesssim0.95H_\mathrm{dz}$ in Fig.\,\ref{fig:tromb_h0025_p1}). This effect can be seen both in high- and low viscosity models. Although this phenomenon is more common in low viscosity models.

As discussed in Section\,\ref{sec:intro}, disc self-gravity tends to decrease the strength and lifetime of vortices in high disc mass models as found by \citep[see e.g.,][]{RegalyandVorobyov2017a}. The stability criterion of the disc against self-gravitational fragmentation can be described by the Toomre $Q$ parameter \citep{Toomre1964} as follows

\begin{equation}
\label{eq:toomre}
    Q = \frac{c_s \Omega}{\pi G \Sigma} > 1.
\end{equation} 

\noindent RWI is triggered in the local minimum of potential vorticity, PV, (referred to as $\zeta$, or vortensity, \citealp[see, e.g.,][]{Lietal2000, Lietal2005, Kolleretal2003}). PV can be expressed as

\begin{equation}
 \label{eq:pv}   
 \mathrm{PV} = \frac{\vec{\omega}}{\Sigma}S^{-2/\gamma},
\end{equation}

\noindent where $\vec{\omega}~=~\nabla\times{\bf{\mathrm{v}}}$ is the vorticity (the curl of the velocity field), and $S~=~P/\Sigma^{\gamma}$ is the entropy. As we assume locally isothermal models, $\gamma=1$, where $\gamma$ is the adiabatic index.

Fig.\,\ref{fig:vort_split} shows the azimuthal distribution of $\delta \Sigma$, PV and the Toomre $Q$ at different stages of the vortex splitting for $h=0.025$, $p=1$, $\Delta r_{\mathrm{dz}} = 0.95H_\mathrm{dz}$ and $M_{\mathrm{d}}/M_{\star} = 0.004$ disc with low viscosity. In Fig.\,\ref{fig:vort_split},  pink lines correspond to the region where the radial averaging was calculated for the normalised density profiles. One can see that PV and $Q$ approach their local minimum value at the same positions where the density contrast reaches its local maximum, which corresponds to the eye of the vortex. With time several local density maxima developed. Since local minima of potential vortensity developed simultaneously, we can conclude that the large-scale vortex split into several small-scale vortices.
 
\begin{figure}
    \centering
    \includegraphics[width=\columnwidth]{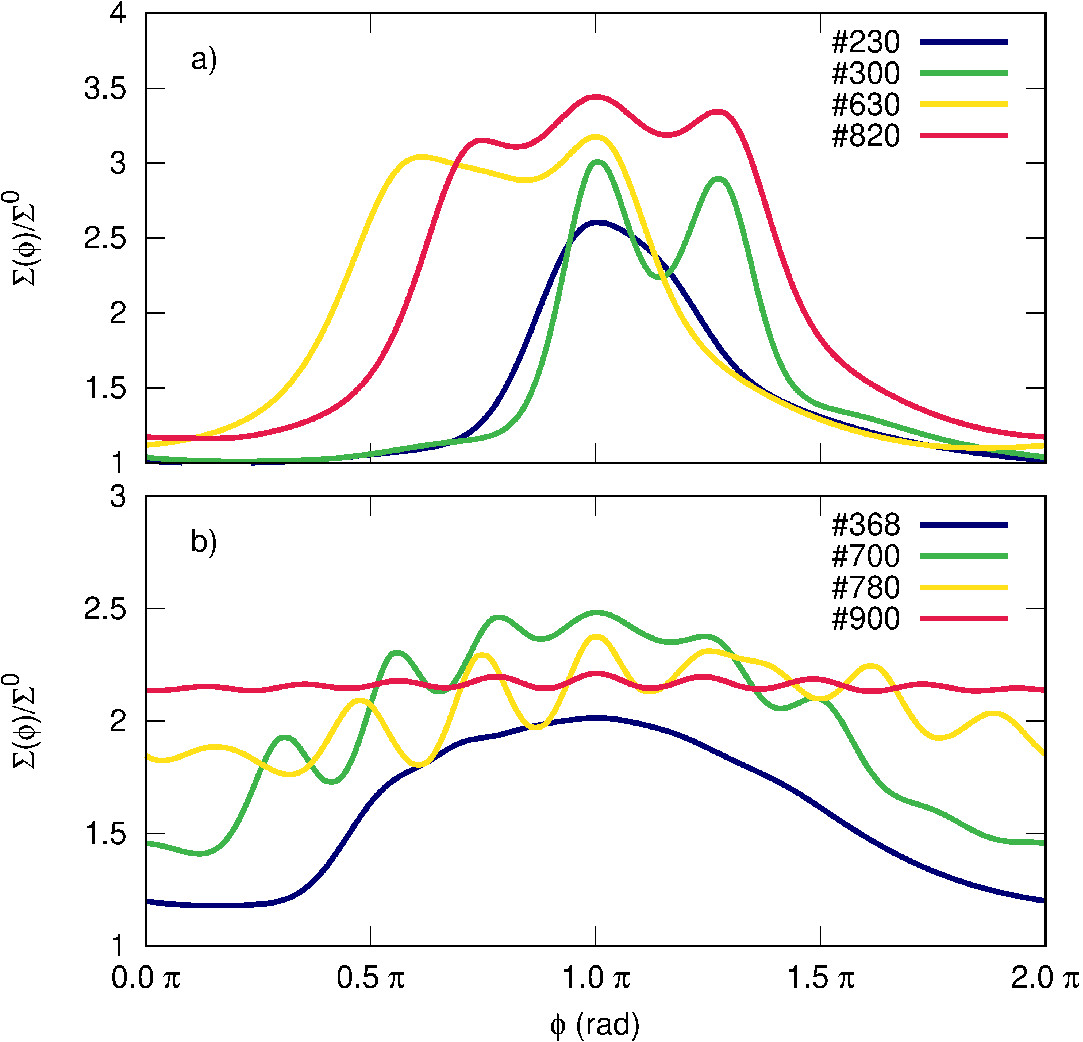}
    \caption{The evolution of $\Sigma/\Sigma^0$ in $h=0.025$, low viscosity ($\alpha_\mathrm{dze}=10^{-5}$), $p=1$ model assuming disc masses of $M_{\mathrm{d}}/M_{\star} = 0.002$ (panel a), and $M_{\mathrm{d}}/M_{\star} = 0.004$ (panel b). In each models $\Delta r_{\mathrm{dz}}$ is assumed to be $0.95H_\mathrm{dz}$. It can be seen that the large-scale vortex splits into a two vortices, which later evolve into a three vortices, see panel a. In a more massive disc (panel b), the large-scale vortex splits into small-scale (more than three) vortices.}
    \label{fig:1d_evol}
\end{figure}

Fig.\,\ref{fig:1d_evol} shows the evolution of the local maximum of the density contrast varies in time. Panels a and b show the evolution of the surface mass density contrast ($\Sigma(\Phi)/\Sigma^0$) in $h=0.025$ model assuming $\alpha_\mathrm{dz}=10^{-5}$. Panels a and b refer to a low disc-mass  of $M_\mathrm{d}/M_\star=0.002$ and $M_\mathrm{d}/M_\star = 0.004$ (which is the same model that is shown in Fig.\,\ref{fig:vort_split})., respectively. The width of the transition region was set to $0.95H_\mathrm{dz}$ in both cases. After a few hundred orbits of evolution a full-fledged large scale vortex 
splits into smaller ones 
(see green, yellow and red lines). 
(blue lines on both panels) tends to 
This effect is common in low-, or middle-disc mass models and low viscosity models. It is appreciable that the azimuthal density contrast of the small-scale vortices decreases with time, see panel b. As a result, vortices dissolve and a gas ring forms. Note that the lifetime of small-scale vortices is longer than our simulation time in the low mass disc model.
 
As discussed in Sec.\,\ref{sec:res}, RWI excitation in $p=1.5$ models was $\Delta r_{\mathrm{dz}}$ limited (beyond a certain width of the viscosity transition region, the disc was RWI stable in the $\Delta r_{\mathrm{dz}} - M_{\mathrm{d}}$ parameter-space). This phenomenon also occurred in $p=1$, $h=0.1$ models. According to equation\,(\ref{eq:toomre}), the Toomre $Q$ is directly proportional to the local speed of sound. This leads to $Q\propto h$ since $c_s = hr\Omega$. Therefore,  increased $h$ stabilises the disc against gravitational fragmentation. Besides, increased $h$ tightens the RWI unstable region in the $\Delta r_{\mathrm{dz}} - M_{\mathrm{d}}$ parameter-space (see Figs.\,\ref{fig:tromb_h0025_p05}, \ref{fig:tromb_h005_p05}, and \ref{fig:tromb_h01_p05}).

\begin{figure*}
    \centering
    \includegraphics[width=\textwidth]{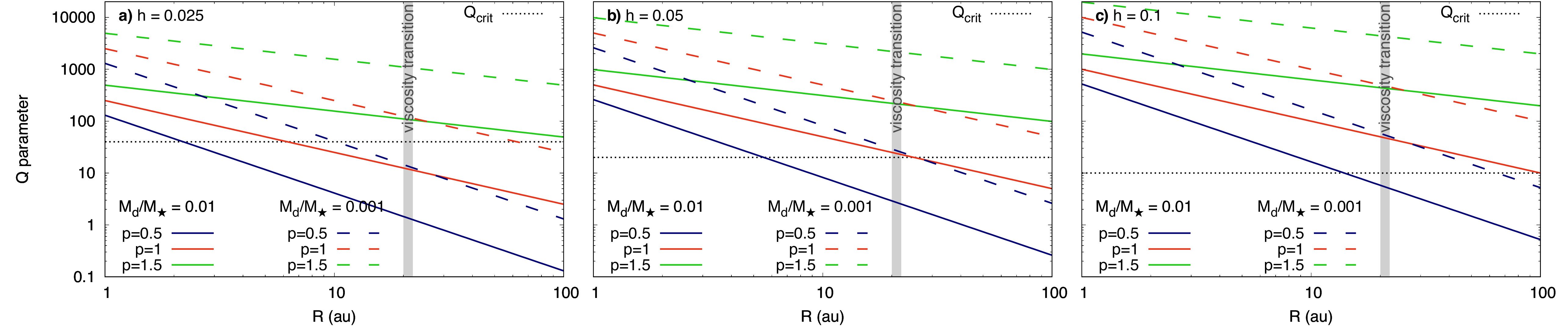}
    \caption{The initial Toomre $Q$ parameter in low ($M_\mathrm{d}=0.001M_\odot$) and high ($M_\mathrm{d}=0.01M_\odot$) mass disc models assuming different $h$ and $p$ parameters. Panel a shows $h=0.025$ cases, while panel b, and c refer to $h=0.05$ and $0.1$ cases. Solid and dashed lines show high and low disc mass models, respectively. Blue, red and green lines correspond to $p=0.5$, $1$ and $1.5$ models. Black dotted lines represent the critical Toomre parameter ($Q_{\mathrm{crit}}$), beyond which no large-scale vortex formation was observed. The grey area represents the dead zone edge where RWI excitation occurred.}
    \label{fig:toomre}
\end{figure*}

Fig.\,\ref{fig:toomre} shows the initial Toomre $Q^0$ parameter for low- and high disc mass models assuming $p=0.5$, $1$, and $1.5$ and the disc aspect ratio as $0.025$, $0.05$, and $0.1$. It can be seen that in both cases where $h$ or $p$ increases, $Q$ also raises. This leads to a gravitationally more stable disc configurations. $Q_{\mathrm{crit}} = 1/h$ is also shown, which defines the critical Q value bellow which the disc self-gravity becomes important with regards vortex evolution according to \citep[see ][]{LovelaceandHohlfeld2013,Yellinetal2016}. Thus, increasing $h$ yields lower $Q_{\mathrm{crit}}$ values.

\begin{table*}
    \caption{Three types of vortex history. Case I represents those models, in which RWI was not excited. In Case II, large-scale vortex formation was occurred ($m=1$), while in Case III, the coagulation of small-scale vortices ($m>1$) was suppressed by the disc self-gravity. Here $m$ refers to the mode number of vortices.}
    \centering
    \begin{tabular}{llccll}
        \hline\hline
         & & RWI excitation & $m$ & disc parameters$^\diamond$&  remark \\
         \hline
         \multicolumn{2}{l}{{\bf Case I}}& \xmark & \nodata & $Q^0<Q_{\mathrm{crit}}$, or large $\Delta r_\mathrm{dz}$&\nodata\\
         \noalign{\smallskip}
         \noalign{\smallskip}
         \multirow{3}{*}{{\bf Case II}}&{\bf a)} & \cmark & $1$ & $Q^0\lesssim Q_{\mathrm{crit}}$ in $h=0.025$, $p=0.5,1$ models  & short-term$^\ast$ vortex\\
         & {\bf b)} & \cmark & $1$ & $Q^0< Q_{\mathrm{crit}}$ in $h=0.025$, $p=0.5$ models  & vortex re-formation\\
         & {\bf c)} & \cmark & $1$ & $Q^0\ll Q_{\mathrm{crit}}$ in $h=0.025$, $p=0.5$ models  & long-term$^\dagger$ vortex\\
         \noalign{\smallskip}
         \noalign{\smallskip}
         \multirow{3}{*}{{\bf Case III}}&{\bf a)} & \cmark & $>1$ & $Q^0\gtrsim Q_{\mathrm{crit}}$ & short-term$^\ast$ vortices\\
         & {\bf b)} & \cmark & $>1$ & $Q^0> Q_{\mathrm{crit}}$ in $\alpha=10^{-5}$ models & vortex re-formation\\
         & {\bf c)} & \cmark & $>1$ & $Q^0\gg Q_{\mathrm{crit}}$ & long-term$^\dagger$ vortices\\\hline\hline
        \multicolumn{6}{l}{$^\diamond${\footnotesize $Q^0$ and $Q_\mathrm{crit}$ are the initial and the critical values of the Toomre parameter at the distance of the vortex eye, see more details in}}\\
        \multicolumn{6}{l}{Section~\ref{sec:vort_split}}\\
        \multicolumn{6}{l}{$^\ast${\footnotesize Vortex dissipates within the simulation time}}\\
        \multicolumn{6}{l}{$^\dagger${\footnotesize Vortex lasts longer than the simulation time}}\\
    \end{tabular}
    \label{tab:vort_cases}
\end{table*}

As shown in Fig. \ref{fig:trom}, three distinct modes of evolution of the pressure jump are found in this study. Table~\ref{tab:vort_cases} summarises the three cases with the corresponding disc parameters. One can see that the different cases can be separated by the initial value of the Toomre parameter. Discs with initially small $Q$ values ($Q^0\leq Q_{\mathrm{crit}}$) cannot sustain large-scale vortices \cite[see, e.g.,][]{RegalyandVorobyov2017a}, however, long-lived small-scale vortices can be formed, e.g. mostly in $h=0.025$, $p=0.5$ models. This phenomenon can be explained by that the disc self-gravity suppresses or even prohibits the coagulation process of small-scale vortices. 

In a more massive disc, if $Q^0\lesssim 1$, the disc fragments into clumps at early stages (a few tenths of orbits). However, in the following phase, small-scale vortices form but are sustained only for a few tenths of orbits. Then small-scale vortices tend to dissolve, forming a ring-shaped gas accumulation. In the subsequent few tenths of orbits, small-scale vortices reappear, leading to a cycle of alternating between small-scale vortices and the ring phases in $h=0.025$, $p=0.5$ middle- and high disc mass models (see, e.g., Fig.\,\ref{fig:frag}). In these cases, the self-gravity of the gas inhibits the coagulation process of small-scale vortices. Therefore large-scale vortex cannot be formed.

If $Q^0 \gtrsim Q_{\mathrm{crit}}$, large-scale vortices can form. However, they tend to split into smaller vortices or elongate and dissipate as they evolve. This phenomenon can be seen in small disc mass models ($M_\mathrm{d}/M_\star \lesssim 0.004$). Therefore it is not a result of the disc's self-gravity. To explore the effect that causes this phenomenon, further investigations are needed.

\begin{figure}
    \centering
    \includegraphics[width=\columnwidth]{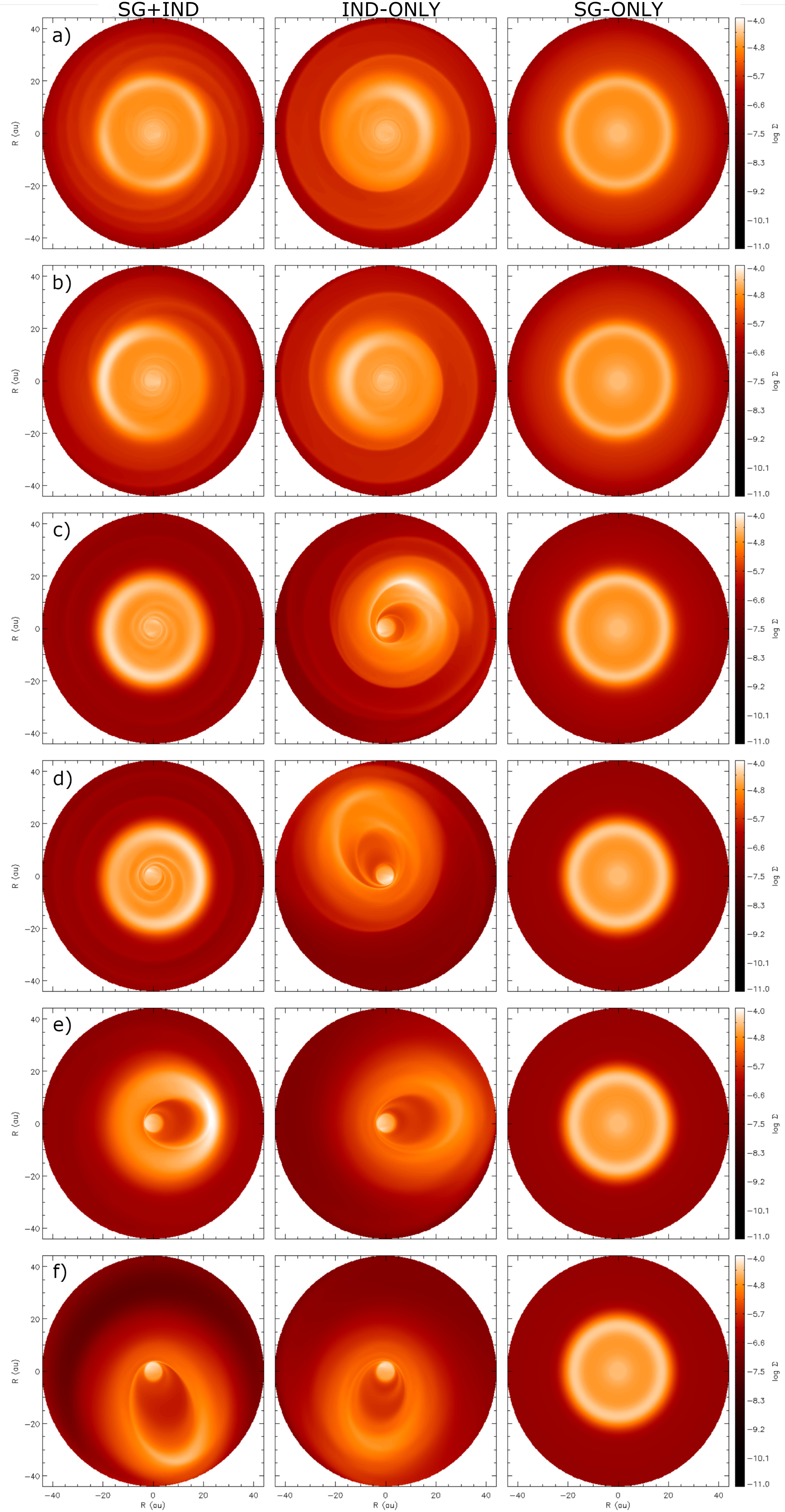}
    \caption{The effect of the indirect potential on the evolution of vortices. Each row refers to the same number of orbits at the distance of the vortex (from top to bottom: 72, 80, 200, 441, 504, and 563). As a comparison, the first column refers to the fiducial case (in which self-gravity and the indirect term are also taken into account, SG+IND model), the second column refers to the case when only the indirect potential is taken into account and self-gravity is neglected (IND-ONLY). The third column represents when the indirect term is neglected, but self-gravity is taken into account (SG-ONLY). In each models, $p$ was set to $0.5$, while $h$ and $\alpha_\mathrm{dz}$ was set to $0.1$, and $10^{-4}$, respectively. The width of the transition region, $\Delta r_\mathrm{dz}$ was set to $0.8H_\mathrm{dz}$ and we assumed that the disc mass is $M_\mathrm{d}=0.01M_\star$.}
    \label{fig:noind_nosg}
\end{figure}

If the growth rate of gas accumulation at the edge of the dead zone is large, the mass of the gas content in the vortex shifts the barycentre of the system, which affects the evolution of the disc as it was shown by \cite{RegalyandVorobyov2017b}. Such effects can be seen, when $p$ is set to $0.5$ and $h$ is $0.1$ or $0.05$. In these cases, the vortex tends to elongate radially due to the effect of the indirect potential of the disc. As the radial extension of the vortex increases, the disc starts to wobble around the barycentre, resulting in a highly eccentric vortex shape. 

Fig.\,\ref{fig:noind_nosg} shows the disc evolution from the orbital period of 70 to the end of the simulation assuming $\alpha=10^{-4}$, $p=0.5$, $h=0.1$, $\Delta r_\mathrm{dz}=0.8H_\mathrm{dz}$, $M_\mathrm{d}=0.01M_\star$. The left-hand-side column shows the fiducial model where both self-gravity and the indirect potential was taken into account (referred to as SG+IND). In order to explore the effect of the barycentre shift of the system, we run two additional test simulations. The middle column of Fig.\,\ref{fig:noind_nosg} represents the evolution of gas with indirect potential (neglecting disc self-gravity, referred to as IND-ONLY). The right-hand-side column shows the evolution of gas with self-gravity (without indirect potential, referred to as SG-ONLY). If the self-gravity is neglected while the indirect potential is taken into account (IND-ONLY model), the disc becomes elliptical inside the dead zone at an earlier stage than in the self-gravitating model (SG+IND model). If the indirect term is not taken into account (SG-ONLY model), the effect of self-gravity prohibits the formation of a large-scale vortex only a ring-shaped gas accumulation form. Therefore, we conclude that the indirect term has a crucial effect on the formation of a large-scale vortex and its lifetime, especially in a disc with a mass of $0.01\,M_\star$. The same conclusion was found by \cite{RegalyandVorobyov2017a}.

\subsection{Effect of viscosity and disc geometry on the vortex strength}

The Navier-Stokes equations (see equation\,(\ref{eq:NS})) that govern the viscous evolution of the gas are dependent on the viscosity. As a result, low viscosity slows down the evolution and lengthens the lifetime of vortices. Our results revealed that decreasing the disc aspect ratio also leads to a longer vortex lifetime. This is in agreement with \cite{TarczayNehezetal2020}. 

We have also shown that large $h$ leads to azimuthally more elongated, therefore weaker vortices. This is the most pronounced in the $p=1.5$ models. 
This phenomenon can be explained by that we used a locally isothermal and flat-disc approximation with $\alpha$-prescription. 
Combining equations (\ref{eq:chshomega}) and (\ref{eq:shakura}) leads to $\nu\sim h^2$, which results that the viscosity of the gas, in some cases, can be lower in $\alpha=10^{-4}$ than in an $\alpha_\mathrm{dz}=10^{-5}$ for sufficiently high values of $h$. Namely, the kinematic viscosity in $h=0.1$, $\alpha=10^{-5}$ models is higher than it is in $h=0.025$, $\alpha=10^{-4}$ ones (see, e.g.. Fig.\,\ref{fig:visc_p}).

As mentioned in Sec.~\ref{sec:res}, the mode number, $m$, depends on the geometric aspect ratio of the disc, $h$. This can be explained by that we assumed locally isothermal $\alpha$ discs. As the kinematic viscosity, $\nu$, depends on $h^2$, one can conclude that higher $h$ values lead to higher viscosities, i.e., smaller mode numbers, see \cite{RegalyandVorobyov2017a}.

\begin{figure}
    \centering
    \includegraphics[width=\columnwidth]{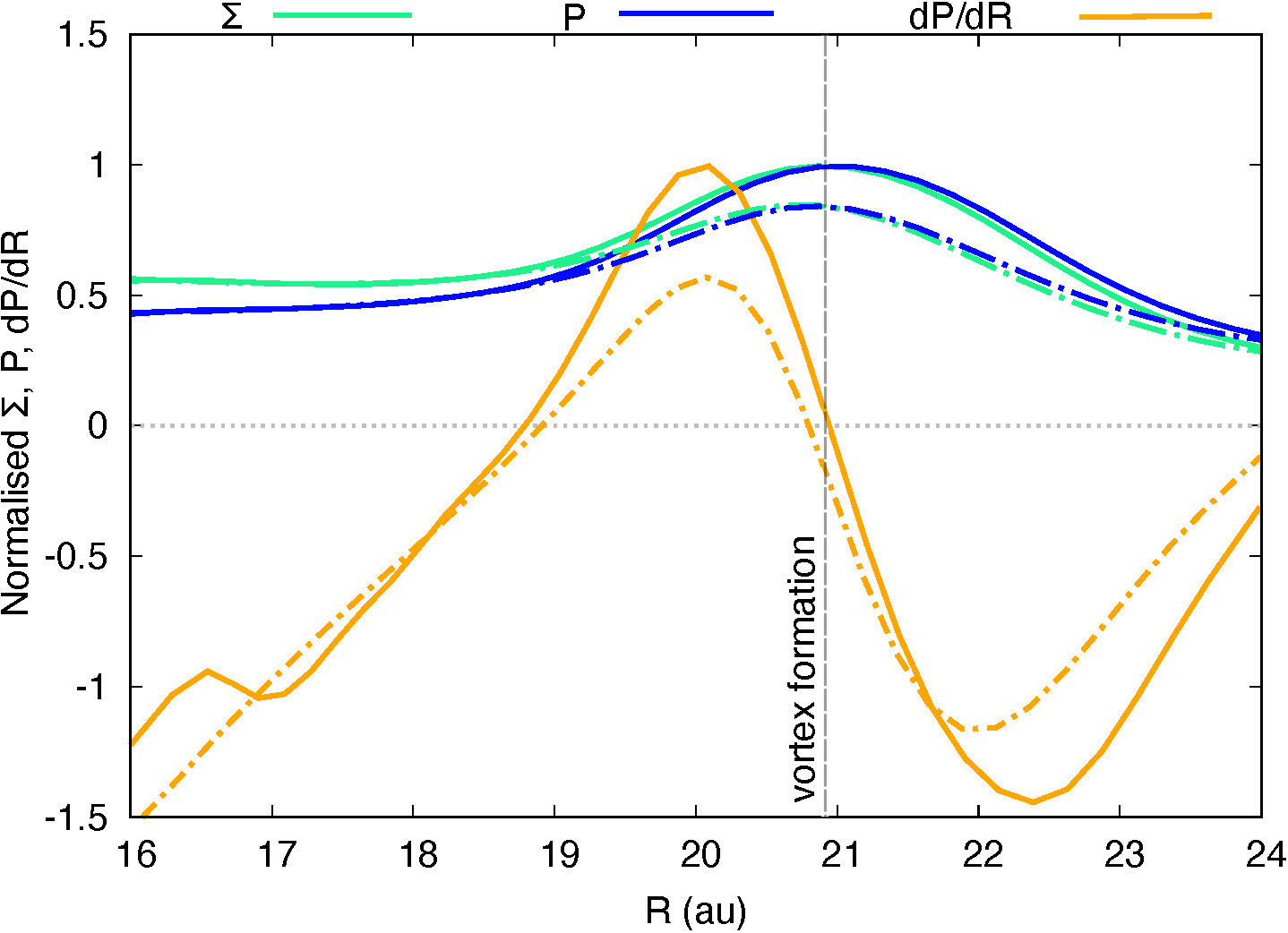}
    \caption{ The azimuthally averaged and normalised radial profile of $\Sigma$ (green lines), gas pressure $P$ (blue lines) and the pressure gradient (yellow lines) in $p=1$, $h=0.05$ low disc mass models ($M_\mathrm{d} = 0.001M_\mathrm{\odot}$). Solid line is the RWI unstable case ($\Delta r_\mathrm{dz}=1.1H_\mathrm{dz}$), while dashed line shows the RWI stable model ($\Delta r_\mathrm{dz}=1.25H_\mathrm{dz}$). Radial profiles for both models are plotted at $t=97$th orbits, when RWI is excited in the RWI unstable model.}
\label{fig:new}
\end{figure}

As it is mentioned in Section~\ref{sec:intro},  RWI is excited at a vortensity minimum of a steep pressure gradient \citep[see][]{Lovelaceetal1999}. The pressure gradient is a function of several parameters such as $h$, $p$, $\nu$, $\Delta r_\mathrm{dz}$ in a locally isothermal $\alpha$ disc, see equations~(\ref{eq:press})-(\ref{eq:chshomega}) and (\ref{eq:sig}). Fig.~\ref{fig:new} shows an RWI unstable ($\Delta r_\mathrm{dz} = 1.1H_\mathrm{dz}$, solid lines) and stable ($\Delta r_\mathrm{dz} = 1.25H_\mathrm{dz}$, dashed lines) low disc mass ($M_\mathrm{d} = 0.001M_\odot$) models for $p=1$, $h=0.05$. 
One can see that the pressure gradient is weaker (and the amplitude of the density bump is smaller) in model where RWI is not excited, than it is in an RWI unstable model. Note that the pressure maximum is somewhat shifted compared to the density maximum due to the locally isothermal approximation, which can be also seen in  \citep[figure 5. in][]{Regalyetal2017}.


\begin{figure}
    \centering
    \includegraphics[width=\columnwidth]{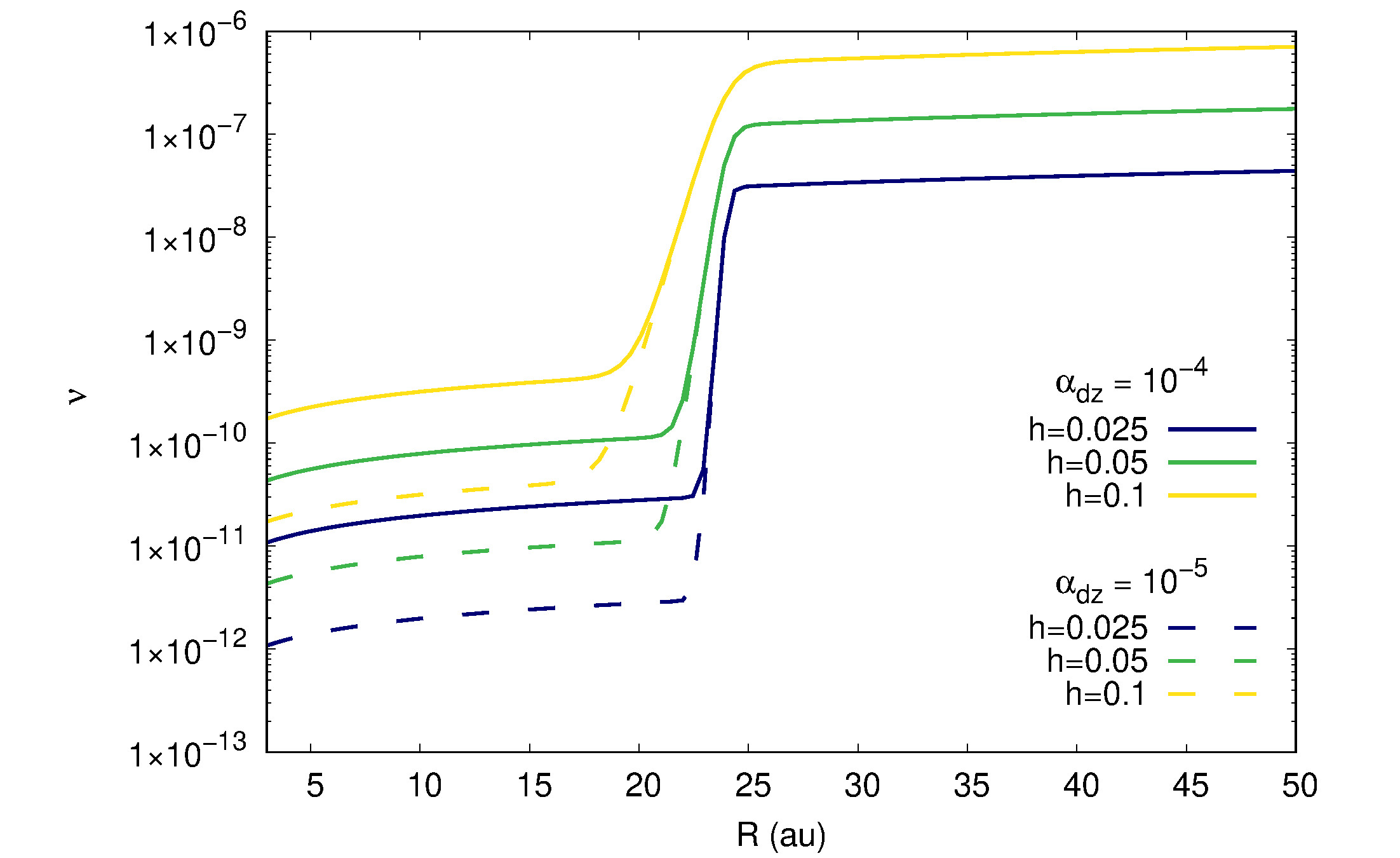}
    \caption{Radial profile of the kinematic viscosity of the disc in $\Delta r_{\mathrm{dz}} = 0.5H_\mathrm{dz}$ models {that assume different disc aspect ratios, $h$}. Solid and dashed lines correspond to $\alpha=10^{-4}$ and $10^{-5}$ models, respectively. It can be seen that disc geometry has an essential effect on the viscosity, e.g. the viscosity in $h=0.025$, $\alpha=10^{-4}$ models is lower in the dead zone than that is in $h=0.1$, $\alpha=10^{-5}$ cases.}
\label{fig:visc_p}
\end{figure}

\subsection{Estimation of a critical disc mass}

In the following, we present our method to estimate an upper limit of the disc mass, for which case a large-scale vortex can be formed assuming a given viscosity transition, disc viscosity and disc geometry. 

\begin{figure*}
    \centering
    \includegraphics[width=\textwidth]{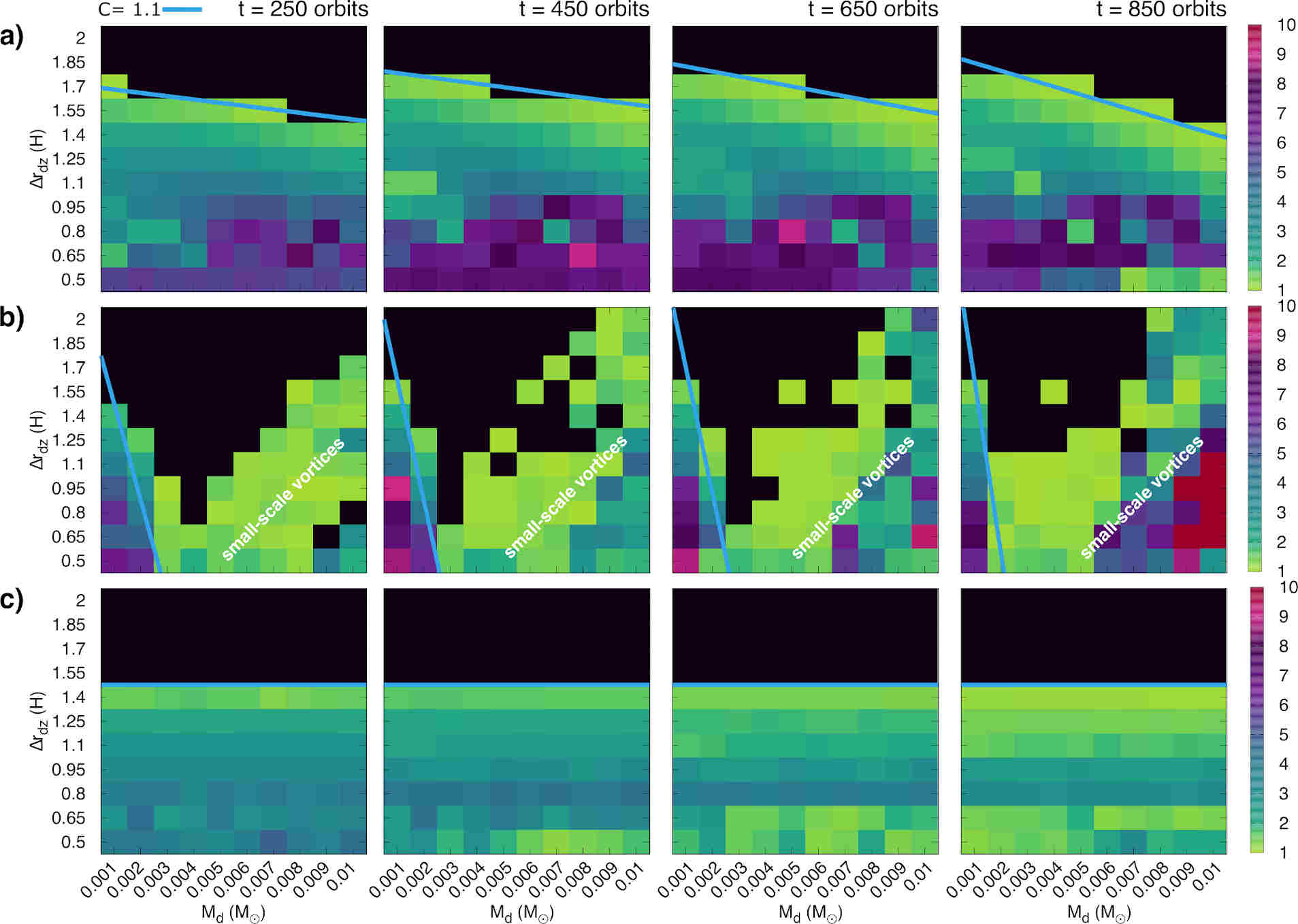}
    \caption{The evolution of the contrast, $C$, as a function of $\Delta r_{\mathrm{dz}}$ and $M_{\mathrm{d}}$. Four different evolutionary stages ($t=250$, $450$, $650$, and $850$ orbits) are shown. Panel a shows $p=1$, $\alpha=10^{-5}$ models. Panel b shows $p=0.5$, $\alpha=10^{-5}$ models, while panel c corresponds to $p=1.5$, $\alpha=10^{-4}$ models. The geometric aspect ratio, $h$, was assumed to be $0.05$ in each models. It can be seen that the $C=1.1$ threshold line (blue line) gets steeper in time, see the time evolution from left to right on panels a and b. On panels b, $C$ is only fitted for large-scale vortices. Panels c shows $\Delta r_{\mathrm{dz}}$ limited cases, thus we were not able to estimate the critical disc mass $M_{\mathrm{crit}}$, see explanation in the text.} 
    \label{fig:heat_map}
\end{figure*}

First, in order to investigate the existence of the vortex, we defined a threshold  ($C$) at $\delta\Sigma \simeq 1.1$. This threshold corresponds to a variation of 10\% in the surface density. Second, we calculated the maximum value of $\delta \Sigma$ for a given model at each time step. Fig.\,\ref{fig:heat_map} shows the maximum value of $\delta \Sigma$ at different evolutionary stages ($t=250$, $450$, $650$, and $850$). The magnitude of the contrast $\delta \Sigma$ is colour coded in each model. Black coloured boxes represent those models in which the contrast did not exceed the critical value of $1.1$ (below which no RWI excitation was observed).  
Fig.\,\ref{fig:heat_map} shows three different $\alpha_\mathrm{dz}$, $h$ and $p$ sets in the $M_\mathrm{dz} - \Delta r_\mathrm{dz}$ parameter field. Panel a and b refer to $h=0.05$, $p=1$ and $p=0.5$ models in the low-viscosity case, respectively. Panel c presents the $h=0.05$, $p=1.5$ models assuming $\alpha_\mathrm{dz}=10^{-4}$. 

Panel a of Fig.\,\ref{fig:heat_map} represents a typical $h - p$ set in which large-scale vortex formation occurred in all RWI unstable models. Panel b represents those cases in which large-scale vortex formation occurred only in small disc mass models. Hence the contrast $C$ was fitted on those models only. Note that for higher disc masses, RWI was excited, however, the coagulation of small-scale vortices was suppressed by the disc self-gravity. Panel c represents those models in which we could not determine the critical disc mass ($\Delta r_\mathrm{dz}$ limited models).

The RWI unstable and stable models are well-separated in most cases. The separation can be determined by a linear regression (see the blue $C=1.1$ line in Fig.\,\ref{fig:heat_map}). Note that the threshold line was determined only for those models where a large-scale vortex formed. Thus, we calculated the intersection of the threshold line and the abscissa ($M_\mathrm{d}$) at each time step. This is an estimated value of the disc mass (referred as $M_\mathrm{est}$) for which case large-scale vortex formation could be triggered by assuming a viscosity transition with infinitesimal width, i.e., $\Delta r_{\mathrm{dz}} = 0H_\mathrm{dz}$. Note that there is a discontinuity in the viscosity in this case, which is fairly unphysical. However, it gives a useful estimation for the critical disc mass.  $M_{\mathrm{est}}$ is an upper limit of the disc mass, for which case a vortex can be maintained at a given time step.

\begin{figure*}
    \centering
    \includegraphics[width=\textwidth]{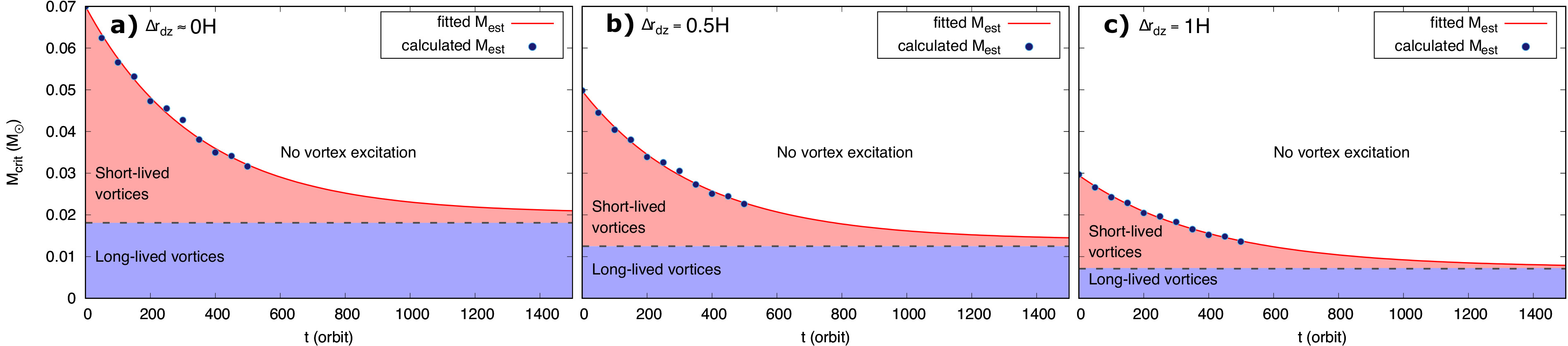}
    \caption{Estimation of $M_{\mathrm{crit}}$ assuming three different viscosity transition width for $h = 0.05$, $\alpha = 10^{-4}$, $p=1$ models. Values for $M_{\mathrm{crit}}$ are calculated in each time-step after the large-scale vortex has been developed. Note that the zero point of $X$-axis corresponds to the phase of the formation of the large-scale vortex. Panel a shows a model, in which the region of viscosity transition is infinitesimally thin, $\Delta r_{\mathrm{dz}} = 0H_\mathrm{dz}$. Panels b, and c present $\Delta r_{\mathrm{dz}} =0.5$ and $1H_\mathrm{dz}$ models. Blue coloured region corresponds to the mass range, in which the lifetime of the large-scale vortex was longer than the simulation time. The dashed line indicates the critical mass, $M_\mathrm{crit}$, while the red area shows the mass range in which the large-scale vortex dissolves within simulation time. No RWI excitation, thus no vortex formation occurred in the white regions.}
    \label{fig:m_crit_p1_h005}
\end{figure*}

The steepness of the threshold line evolves with time, hence  $M_\mathrm{est}$ depends on time. The time evolution of $M_\mathrm{est}$ follows an exponential decay to a minimum (saturation) value as

\begin{equation}
\label{eq:est}
    M_{\mathrm{est}}(t) = c_1 + \frac{c_2}{1+e^{c_3\cdot t}},
\end{equation}

\noindent where $c_1$ is the saturation value, $c_1 + c_2/2$ equals to $M_{\mathrm{est}} (t=0)$ and $c_3$ is a characteristic time for the lifetime of the vortex. In the limit of $t \rightarrow \infty$, equation\,(\ref{eq:est}) gives a critical value for the disc mass ($M_\mathrm{crit}$), below which large-scale vortex can be developed.

To calculate $M_{\mathrm{crit}}$, first, we determined $t^{\mathrm{m=1}}$ phase, where only a large-scale ($m=1$ mode) vortex is present in the disc. Afterwards, we used a non-linear least squares (NLLS) Marquardt-Levenberg algorithm to calculate the $c_1$, $c_2$ and $c_3$ parameters (see e.g. Fig.\,\ref{fig:m_crit_p1_h005}). Summary of the results are shown in Table\,\ref{tab:masses} listing the calculated $M_\mathrm{crit}$ values (and the corresponding $Q^0$ values at the distance of the vortex eye) for different model parameters ($p$, $h$, $\alpha$ and $\Delta r_{\mathrm{dz}}$).

\begin{table*}
\caption{Estimated critical mass for large-scale vortex formation with the corresponding $Q^0$ values at the distance of the vortex eye.}
\label{tab:masses}
\begin{tabularx}{\textwidth}{lcYcYcYcYcYcY}
\hline
& \multicolumn{4}{c}{$p=0.5$} & \multicolumn{8}{c}{$p=1$}\\
\cmidrule(lr){2-5}\cmidrule(lr){6-13}
\multirow{2}{*}{$\Delta r_{\mathrm{dz}}$}  &  \multicolumn{4}{c}{$M_{\mathrm{crit, h=0.05}}$ [$M_\odot$] }   & \multicolumn{4}{c}{$M_{\mathrm{crit, h=0.025}}$ [$M_\odot$]}   & \multicolumn{4}{c}{$M_{\mathrm{crit, h=0.05}}$ [$M_\odot$]}  \\
\cmidrule(lr){2-5}\cmidrule(lr){6-9}\cmidrule{10-13}
$\left[ \mathrm{H}\right]$ &  $\alpha_\mathrm{dz} = 10^{-4}$ & $Q^0$ &  $\alpha_\mathrm{dz} = 10^{-5}$ & $Q^0$ &  $\alpha_\mathrm{dz} = 10^{-4}$ & $Q^0$  &  $\alpha_\mathrm{dz} = 10^{-5}$ & $Q^0$ & $\alpha_\mathrm{dz} = 10^{-4}$ & $Q^0$ & $\alpha_\mathrm{dz} = 10^{-5}$ &$Q^0$  \\

\hline\noalign{\smallskip}
\noalign{\smallskip} 
0   & 0.0019 & 15.32  &    0.0016  & 18.19
& 0.0020 & 62.50 & 0.0018 & 69.44 &  0.0179  & 13.97  &  0.0176 & 14.21\\
\noalign{\smallskip} 
\noalign{\smallskip} 
0.5  &  0.0013 & 22.39  & 0.0011 & 26.46 & 0.0018 & 69.44 & 0.0015 & 83.33 & 0.0128  & 11.64 & 0.0125  & 20.00   \\
0.65  & 0.0011 & 26.46 & 0.0010 & 29.10 & 0.0017 & 73.53 & 0.0014 & 89.29 &   0.0104 & 24.04 & 0.0109 & 22.93    \\
0.8   & 0.0008 & 36.38 & 0.0009 & 32.34 & 0.0015 & 83.33 & 0.0013 & 96.15 & 0.0079  & 31.65  & 0.0099 & 25.25\\
0.95   & 0.0006 & 48.50 & 0.0007  & 41.57 & 0.0014 & 89.29 & 0.0012 & 104.17 &   0.0053  & 47.17  & 0.0079 & 31.65 \\

1.1       & 0.0004 & 72.76 & 0.0006 & 48.50 & 0.0012 & 104.17 & 0.0011 &113.64 & 0.0025  & 100.00 & 0.0059  & 42.37\\
1.25   & 0.0001 & 291.02 & 0.0005 & 58.76 & 0.0010 & 125.00 & 0.0010 & 125.00 &   \xmark &   \xmark  & 0.0038  & 65.79 \\
1.4    & \xmark  & \xmark & 0.0004 & 72.76 &  0.0008 & 156.25 & 0.0009 & 138.89 &  \xmark &   \xmark  & 0.0010 & 250.00  \\
1.55     &  \xmark  & \xmark & 0.0003 & 97.07 & \xmark & \xmark & 0.0007 & 178.57 &  \xmark & \xmark  & \xmark &  \xmark \\
1.7   &  \xmark  & \xmark & 0.0002 & 145.51 &\xmark & \xmark & 0.0006 & 208.33 &   \xmark  & \xmark &   \xmark&   \xmark \\
1.85 & \xmark  & \xmark &  \xmark  & \xmark  & \xmark & \xmark & 0.0004 & 312.50 & \xmark  & \xmark &   \xmark&   \xmark\\
2       & \xmark  & \xmark &  \xmark  & \xmark &\xmark & \xmark & 0.0002 & 625.00 & \xmark  & \xmark &   \xmark&   \xmark\\
\hline\hline
\end{tabularx}
\end{table*}

The critical disc mass, $M_\mathrm{crit}$, depends on the steepness of the initial surface density ($p$), the viscosity ($\alpha_\mathrm{dz}$), the width of the viscosity transition ($\Delta r_\mathrm{dz}$), and the geometric aspect ratio of the disc ($h$).
Comparing the second and eighth columns of Table\,\ref{tab:masses}, it can be seen that  $M_\mathrm{crit}$ was higher in $p=1$ cases than it was in $p=0.5$ model. This can be explained by that $M_\mathrm{crit}$ is lower in in $p=0.5$, as was discussed in the previous section.


In general, $M_\mathrm{crit}$ is higher in high viscosity models than it is in low viscosity models. Increasing $\Delta r_\mathrm{dz}$ or $h$ causes lower $M_\mathrm{crit}$. Comparing $h=0.025$ and $0.05$ models in $p=1$ case we found that $M_\mathrm{crit}$ is about ten times higher in $h=0.05$ models than it is in $h=0.025$ models. 

Assuming an infinitesimally small viscosity transition ($\Delta r_\mathrm{dz} = 0H_\mathrm{dz}$), the critical disc mass reaches the lowest value ($M_\mathrm{crit} = 0.0016M_\odot$) in $\alpha_\mathrm{dz} = 10^{-5}$, $p=0.5$, $h=0.05$ case. While the highest value for $M_\mathrm{crit}$ was determined in the high viscosity model, $\alpha_\mathrm{dz} = 10^{-4}$, $p=1$, $h=0.05$. In this case, $M_\mathrm{crit}$ is more than ten times higher (i.e. $M_\mathrm{crit}=0.0179M_\odot$) than it is in the case of the lowest value.

According to  \cite{LovelaceandHohlfeld2013,Yellinetal2016}, the self-gravity is important in discs with assuming $Q_\mathrm{crit} = 1/h$. In discs assuming $h=0.025$, $Q_\mathrm{crit} = 40$, while in $h=0.05$ discs $Q_\mathrm{crit} = 20$. Table\,\ref{tab:masses} contains the corresponding $Q^0$ values for the calculated $M_\mathrm{crit}$ values at the distance of the vortex eye. It can be seen that, assuming sharp viscosity transitions, the corresponding $Q^0$ values for the estimated $M_\mathrm{crit}$ values are close to $Q_\mathrm{crit}$. In the case of an infinitesimally small transition ($\Delta r_\mathrm{dz} = 0H_\mathrm{dz}$) in $h=0.05$ models $Q^0\sim 14-18$. In $h=0.025$ models $Q^0\sim60-70$ (assuming $\Delta r_\mathrm{dz} = 0H_\mathrm{dz}$), which is higher than $Q_\mathrm{crit}$. This is consistent with the previous work of \cite{RegalyandVorobyov2017a}, who found that self-gravity could affect the formation of large-scale vortices in $Q^0>Q_\mathrm{crit}$ discs (e.g. $Q^0=50$ in a canonical disc, assuming $h=0.05$).

We could not determine the critical disc mass in the following cases. RWI excitation in $p=1.5$ models are $\Delta r_\mathrm{dz}$ limited in the $\Delta r_\mathrm{dz} - M_\mathrm{d}$ rather than disc mas limited (in the investigated mass range). In these models, the disc is stable against gravitational fragmentation, e.g. $Q^0 = 220.89$ even for the highest disc mass, assuming $h=0.05$. Note that in $p=1.5$ models, $Q$ remains high throughout the simulation. This means that disc self-gravity cannot suppress large-scale vortex formation with the investigated disc mass range. Hence, $M_\mathrm{crit}$ is far beyond the investigated disc range in these models, which is beyond the scope of this investigation. 

The coagulation of small-scale vortices did not occur in $p=0.5$, $h=0.025$ models. Moreover, in $p=0.5$, $h=0.1$ models, the disc becomes highly eccentric. This can be explained by that the mass of the accumulated gas becomes sufficiently high that the barycentre of the star-disc system shifts. Due to the indirect potential, the disc tends to wobble around the barycentre, which leads to a highly eccentric disc, which is finally disrupted within the time span of our simulation. The effect of self-gravity and the indirect potential on the long-term evolution of the large-scale vortex and the disruption of the disc is shown in Fig.\,\ref{fig:frag}.

\section{Conclusion}
\label{sec:concl}

In this study, we investigated the long-term evolution of vortices formed via RWI excitation at the outer edge of the dead zone in locally isothermal, self-gravitating protoplanetary discs. We performed 1980 2D hydrodynamic simulations in high- and low viscosity regimes ($\alpha_{\mathrm{dz}}=10^{-4}$ and $10^{-5}$). We investigated the effect of disc mass being in the range of $M_{\mathrm{d}}/M_{\star} = 0.01-0.001$. We run models assuming  three different disc aspect ratios ($h=0.025$, $0.05$, and $0.1$ in flat disc approximation), three initial surface density slopes  ($p=0.5$, $1$ and $1.5$). The width of the viscosity transition region was assumed to be in the range of $0.5 - 2H_{\mathrm{dz}}$.  All investigated disc parameters  are listed in Table\,\ref{tab:inits}. In our models, the inner and outer boundaries of the disc were set to $3$ and $50$AU with a resolution of $256$ logarithmic radial and $512$ azimuthal grid cells. 

Based on the simulations, we estimated a critical disc mass, below which the formation of large-scale vortices is allowed. Beyond this critical disc mass, the effect of disc self-gravity slows or even suppresses the coagulation process of small-scale vortices, hence prevents the formation of a large-scale vortex. Table\,\ref{tab:masses} shows the critical values for the disc mass (and the corresponding initial $Q^0$ values) in models where it could be derived. Our main findings are the followings:

1) In low viscosity models, RWI excitation can be triggered at wider viscosity transition regions than in high viscosity models. Comparing to high viscosity models, the critical $\Delta r_\mathrm{dz}$ below which RWI can be excited is $0.3H_{\mathrm{dz}}$ wider in low viscosity models if the disc aspect ratio is $h=0.025$ or $0.05$. However, for $h=0.1$ models, the critical transition width is $0.15H_{\mathrm{dz}}$ wider in low viscosity models.

2) Low $h$ (hence low viscosity) and low $p$ values tend to increase the initial mode number, $m$, and slow down the coagulation of small-scale vortices. Contrary, increasing $h$ and $p$ enhance the formation of large-scale vortices. 

3) Vortex oscillation (break up followed by re-formation of the vortex) is common in low viscosity $h=0.05$ models assuming $p=1$ or $1.5$. In these cases, the lifetime of the reappeared large-scale vortex larger than the simulation time. 

4) Vortex splitting (large-scale vortices tend to break up into smaller ones) frequently occurs in $p=1$, $h=0.025$ models assuming low disc masses.

5) The critical disc mass, below which large-scale vortex formation occurred, was calculated for three different $h-p$ sets: i) in low disc mass models in $h=0.05$, $p=0.5$, ii) $h=0.025$, $p=1$, and iii) $h=0.05$, $p=1$ models (see Table\,\ref{tab:masses}). In general, the critical disc mass is found to be in the order of $0.0016 - 0.012M_\odot$. The corresponding $Q^0$ values are $\sim 15-70$, which depends on $\Delta r_\mathrm{dz}$, $h$, $p$. For an infinitesimally thin viscosity transition ($\Delta r_\mathrm{dz} = 0H_\mathrm{dz}$), $Q^0 \gtrsim Q_\mathrm{crit}$ at the distance of the vortex eye, where $Q_\mathrm{crit} = 1/h$. This is consistent with the previous works of \cite{LovelaceandHohlfeld2013, Yellinetal2016, RegalyandVorobyov2017a}.

6) Ring-like structures develop in the gas for $h=0.025$, $p=0.5$ models, assuming high disc masses ($M_\mathrm{d}/M_\star \gtrsim 0.006 M_\odot$), which resemble the structures that were found by the DSHARP project \cite[see e.g.][]{Dullemondetal2018}. The ring becomes RWI unstable after a few tenths or hundreds of orbits. However, due to the disc self-gravity, the coagulation of small-scale vortices are prevented. At later epochs, an oscillation between the ring-like structure and small-scale vortices was found (see, e.g.,  Fig.\,\ref{fig:frag}).

Here we have to mention some caveats of our models whose resolution requires further investigations. We assumed a locally isothermal disc approximation. This assumption can provide a good approximation as long as the thermal heating and cooling processes are rapid. \cite{PierensandLin2018} and \cite{TarczayNehezetal2020} showed that disc thermodynamics affect vortex strength and lifetime. Thus, in order to explore vortex evolution, the lifetime with different physical parameters, and the critical mass of the disc, further investigations, including disc thermodynamics, are needed. 

In order to investigate the effect of the width of the transition region at the edge of the outer dead zone on the evolution of large-scale vortices, we used a static model of the dead zone (i.e., $\alpha$-prescription with a fixed distance of the transition region). This way, in the locally isothermal approximation, the viscosity depends only on the distance and the $\alpha$-parameter. However, in a more realistic model, the viscosity of the gas is dependent on the surface density of the gas. To investigate the effect of a surface density dependence of the viscosity of the gas, this effect is also needed to be included in our model for further investigations.

We could not estimate $M_\mathrm{crit}$ in those models in which large-scale vortex formation not occurs. $M_\mathrm{crit}$ could not be calculated either in those models, in which the disc becomes eccentric, or $M_\mathrm{crit}$ is far beyond the investigated disc mass range. The latter occurs in models, where the Toomre $Q$ remains high during the whole simulation ($p=1.5$ models). Hence, to estimate $M_\mathrm{crit}$ in $p=1.5$ models, the investigated disc mass range needs to be extended in a future study. Note that, according to \cite{Onoetal2016}, the threshold of 10\% variation of the surface density with respect to the initial one gives us an upper bound estimate of $M_\mathrm{crit}$. To give a more realistic restriction of the critical disc mass, investigating the circumstances of the excitation of RWI would be needed in a further study, i.e., defining the amplitude of the critical surface density variation, that is required to the excitation of RWI.

We used two-dimensional, thin disc approximation, while the theoretical work of \cite{LesurandPaploizou2009} revealed vortex formation in three dimensions face the problem of elliptical instability, which can destroy vortices with $\chi_\mathrm{dens}<4$. Hence, to investigate the effect of different disc parameters on the critical disc mass, further investigations in three dimensions are needed. Note, however, that disc self-gravity implies enormous computational difficulties in three dimension.

To summary, we conclude that long-lived, large-scale vortex formation, at the outer edge of the dead zone, favours $Q^0$ to be orders of magnitudes higher than $Q_\mathrm{crit}$. In $p=0.5$, $h=0.05$ and $p=1$, $h=0.025$ models, the critical disc mass is $\sim0.0016-0.002M_\odot$. Above this disc mass, large-scale vortex formation is suppressed by the self-gravity of the disc. In $p=1$, $h=0.05$ models, the critical disc mass is $\sim10$ times larger than in the previous two cases ($p=0.5$, $h=0.05$ and $p=1$, $h=0.025$ models), see Table\,\ref{tab:masses}. Although, RWI can be excited if the initial value of $Q$ is close to $Q_\mathrm{crit}$, large-scale vortices break up into small ones in $h=0.025$, $p=1$ models for $M_\mathrm{d}/M_\star \lesssim 0.004M_\odot$ disc-masses. If the initial $Q$ is below $Q_\mathrm{crit}$, only small-scale vortices form in an RWI unstable disc, which cannot be merged into one single vortex (e.g. in $h=0.025$, $p=1$ models for $M_\mathrm{d}/M_\star\lesssim0.003M_\odot$ disc-masses. We conclude that long-lived, large scale-vortex formation, therefore a hypothetical vortex-aided formation of planets, favours discs with $Q^0\gg Q_\mathrm{crit}$. Moreover, large-scale vortex formation favours low disc masses and low kinematic viscosity, which conditions are fulfilled in transition discs. Thus, the presence of a large-scale vortex could be an indication to the lifetime of the disc, i.e., they might be more common in transition discs \citep[see, e.g.,][]{Regalyetal2012}.

\section*{Acknowledgements}

This project was funded by the OTKA-119993 grant. DTN acknowledges the support by the Lend\"ulet Program of the Hungarian Academy of Sciences, project No. LP2018-7/2019.
We gratefully acknowledge the support of the NVIDIA Corporation with the donation of the Tesla 2075 and K40 GPUs. We acknowledge KIF\"U for awarding us access to a resource based in Hungary.
This research was supported by the Australian Research Council Centre of Excellence for All Sky Astrophysics in 3 Dimensions (ASTRO 3D), through project number CE170100013. DTN acknowledges the support of the MW-Gaia COST Action (CA 18104) grants and the KKP-137523 'SeismoLab' \'Elvonal grant of the Hungarian Research, Development and Innovation Office (NKFIH). DTN acknowledges E. Vorobyov for his helpful remarks on the manuscript. We thank the anonymous referee for his/her useful comments and remarks.


\section*{Data Availability}

We provide additional figures presenting the evolution of $\Delta \Sigma$ for $\alpha_\mathrm{dze}=10^{-5}$, $h$, $p$ data-sets, similar to Fig.\,\ref{fig:tromb_h005_p1}, in the online supplementary material.





\bibliographystyle{mnras}
\bibliography{main} 



\appendix

\section{Evolution of \texorpdfstring{$\delta \Sigma$}{dS}}
\label{sec:app}
\begin{figure*}
    \centering
    \includegraphics[width=0.85\textwidth]{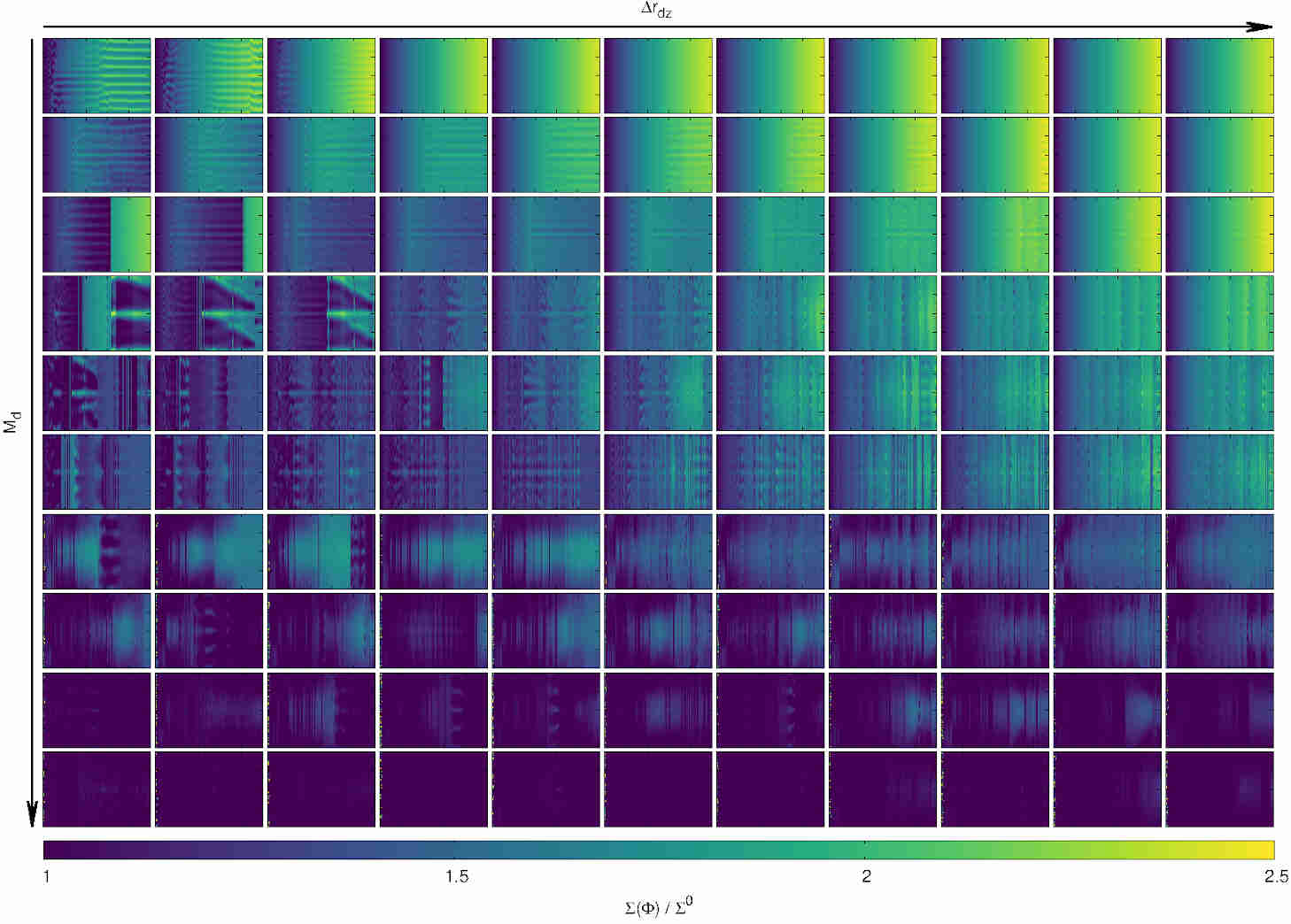}
    \caption{Evolution of $\delta \Sigma$ in $h = 0.025$, $p=0.5$, $\alpha = 10^{-4}$ models. The width of the transition region, and the disc mass increase from left to right and top to bottom, respectively.}
    \label{fig:tromb_h0025_p05}
\end{figure*}

\begin{figure*}
    \centering
    \includegraphics[width=0.82\textwidth]{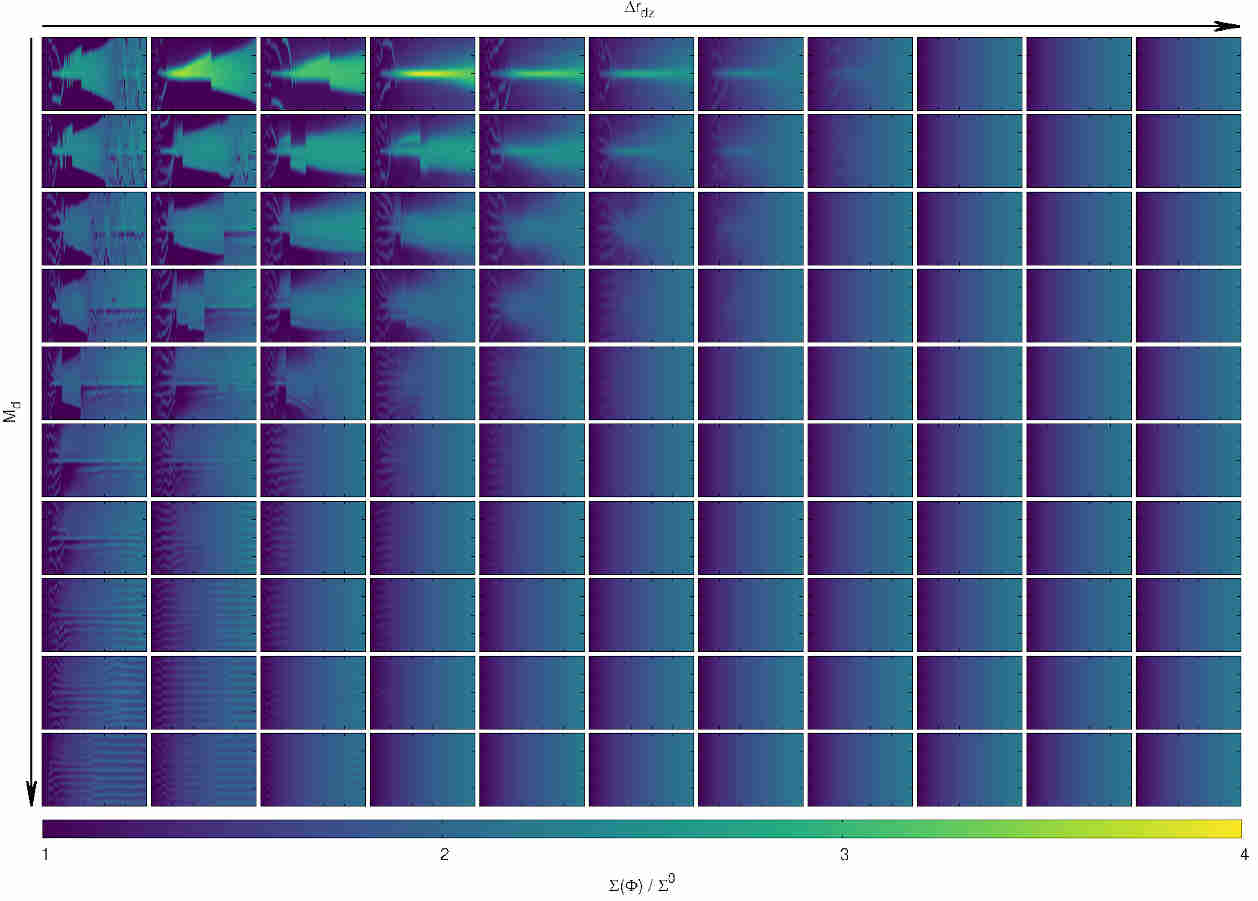}
    \caption{Evolution of $\delta \Sigma$ in $h = 0.025$, $p=1$, $\alpha = 10^{-4}$ models.  The width of the transition region widens, and the disc mass increases from left to right and from top to bottom, respectively.}
    \label{fig:tromb_h0025_p1}
\end{figure*}

\begin{figure*}
    \centering
    \includegraphics[width=0.82\textwidth]{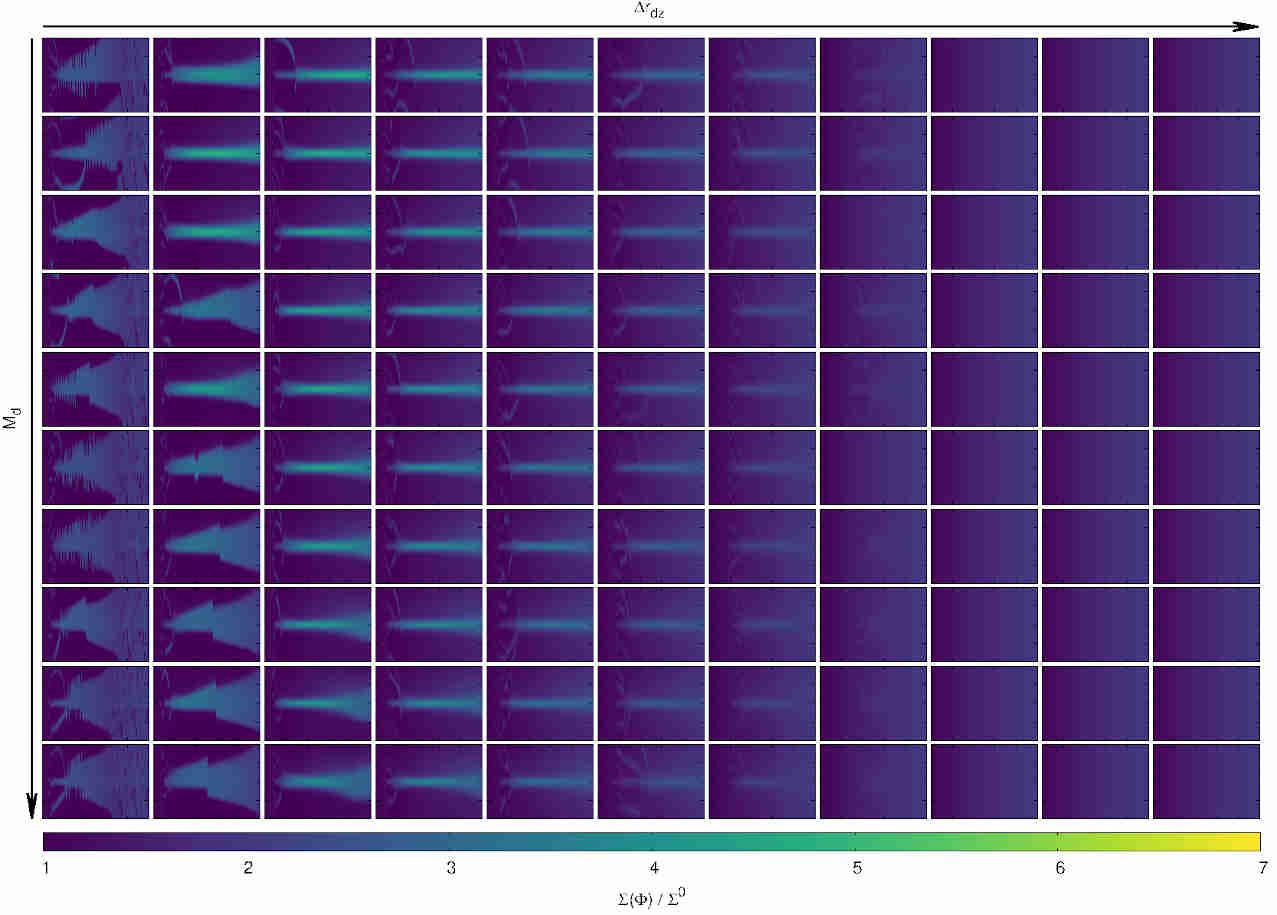}
    \caption{Evolution of $\delta \Sigma$ in $h = 0.025$, $p=1.5$, $\alpha = 10^{-4}$ models.  The width of the transition region widens, and the disc mass increases from left to right and from top to bottom, respectively.}
    \label{fig:tromb_h0025_p15}
\end{figure*}

\begin{figure*}
    \centering
    \includegraphics[width=0.82\textwidth]{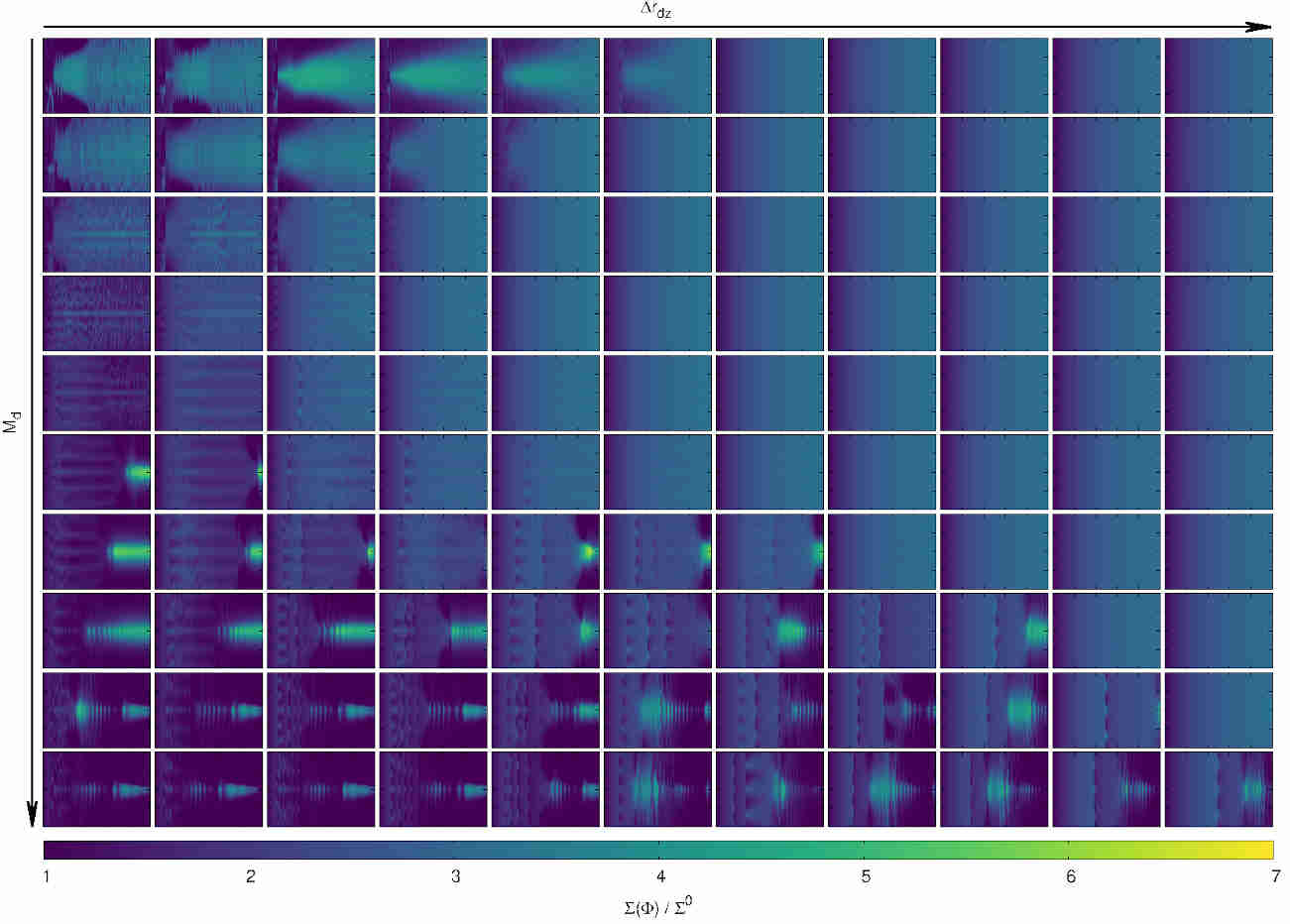}
    \caption{Evolution of $\delta \Sigma$ in $h = 0.05$, $p=0.5$ models, $\alpha = 10^{-4}$ models.  The width of the transition region widens, and the disc mass increases from left to right and from top to bottom, respectively.}
    \label{fig:tromb_h005_p05}
\end{figure*}

\begin{figure*}
    \centering
    \includegraphics[width=0.82\textwidth]{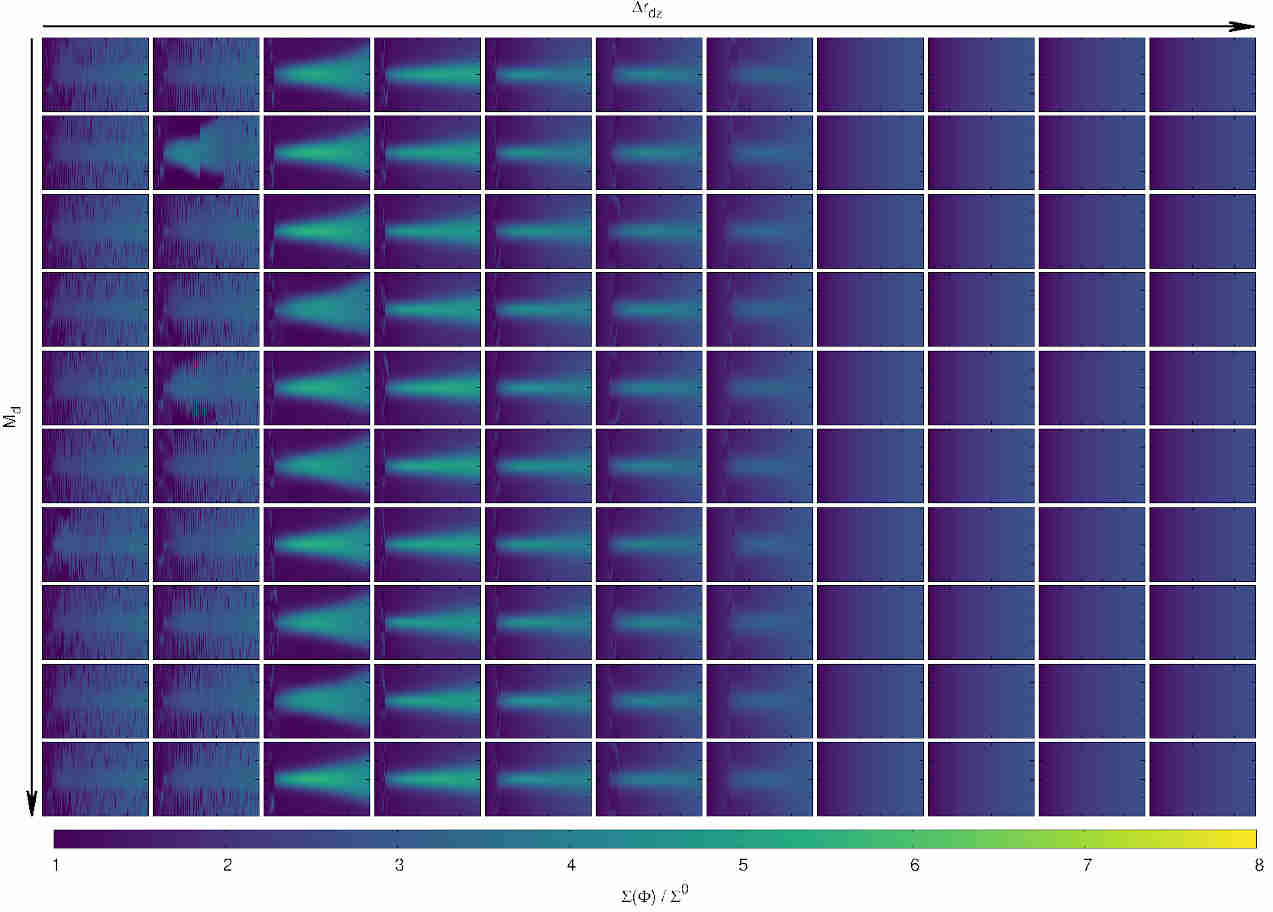}
    \caption{Evolution of $\delta \Sigma$ in $h = 0.05$, $p=1.5$, $\alpha = 10^{-4}$ models.  The width of the transition region widens, and the disc mass increases from left to right and from top to bottom, respectively.}
    \label{fig:tromb_h005_p15}
\end{figure*}

\begin{figure*}
    \centering
    \includegraphics[width=0.82\textwidth]{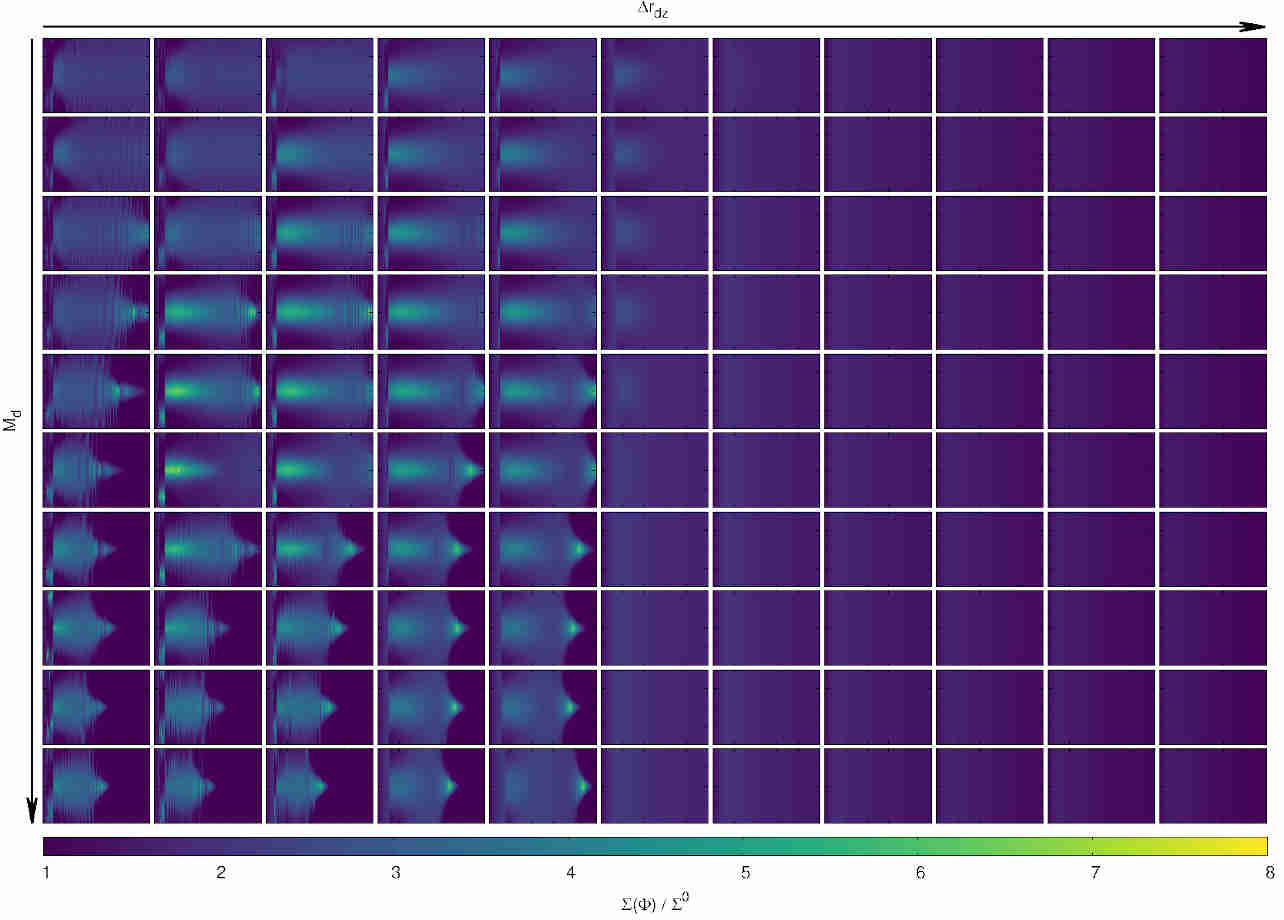}
    \caption{Evolution of $\delta \Sigma$ in $h = 0.1$, $p=0.5$, $\alpha = 10^{-4}$ models.  The width of the transition region widens, and the disc mass increases from left to right and from top to bottom, respectively.}
    \label{fig:tromb_h01_p05}
\end{figure*}

\begin{figure*}
    \centering
    \includegraphics[width=0.82\textwidth]{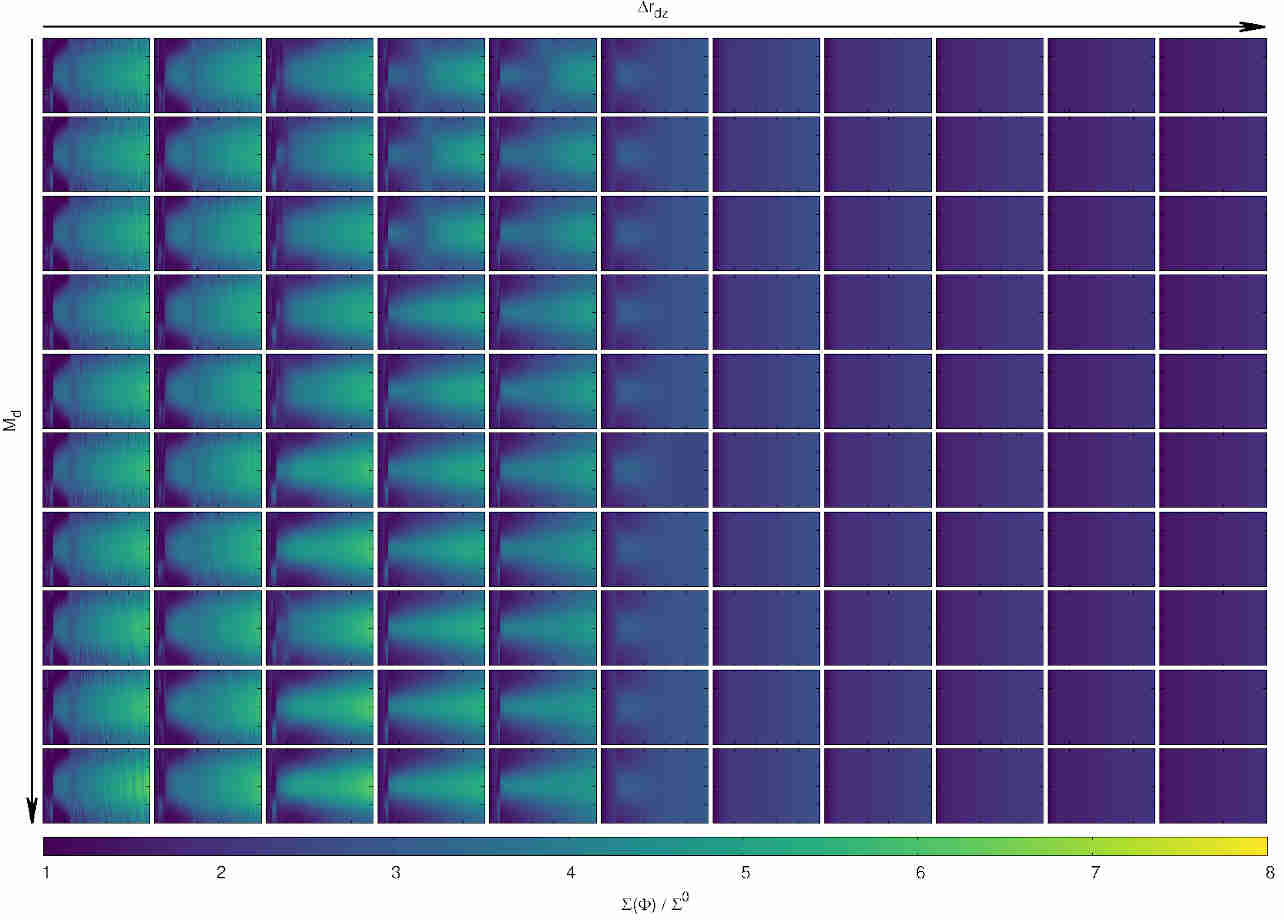}
    \caption{Evolution of $\delta \Sigma$ in $h = 0.1$, $p=1$, $\alpha = 10^{-4}$ models.  The width of the transition region widens, and the disc mass increases from left to right and from top to bottom, respectively.}
    \label{fig:tromb_h01_p1}
\end{figure*}

\begin{figure*}
    \centering
    \includegraphics[width=0.82\textwidth]{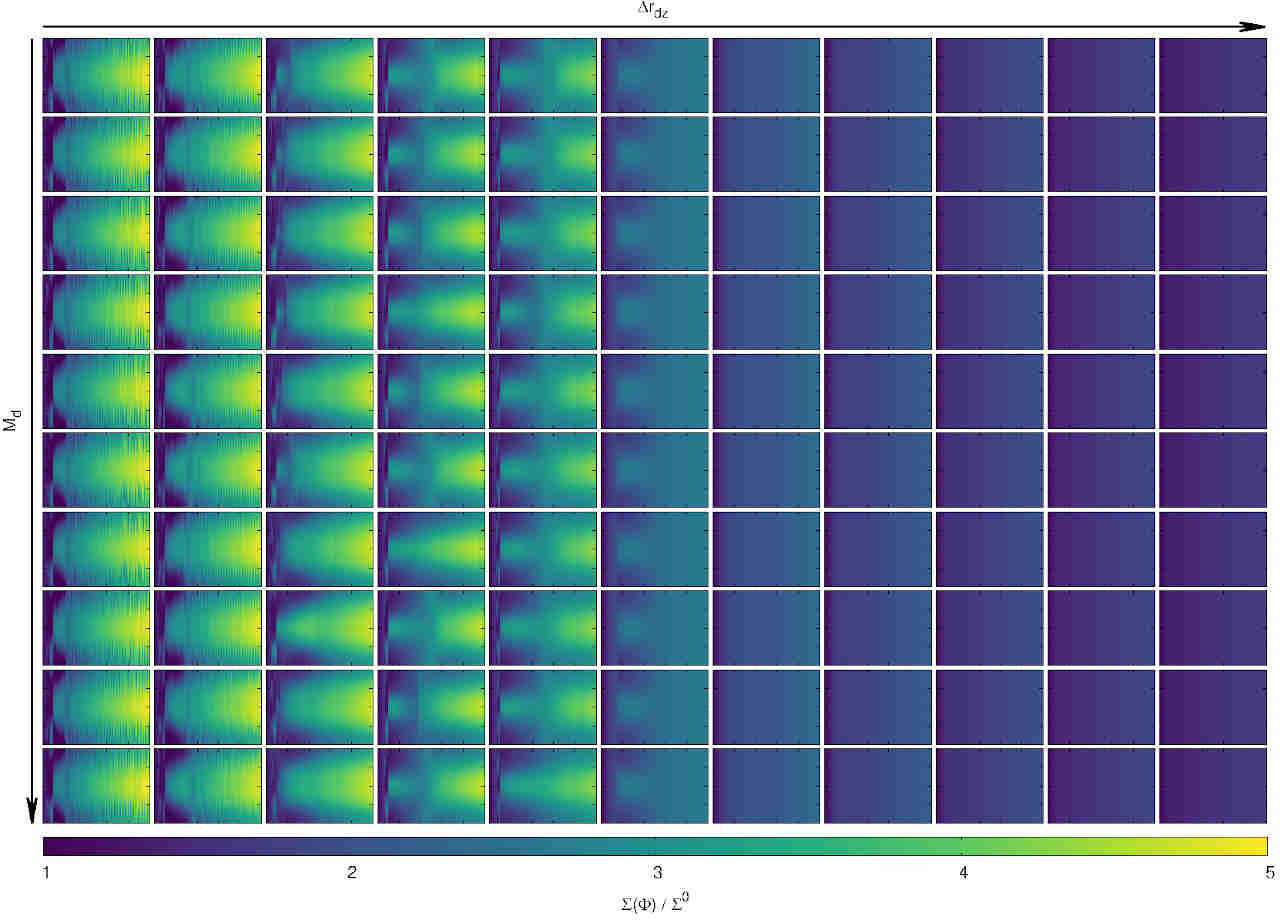}
    \caption{Evolution of $\delta \Sigma$ in $h = 0.1$, $p=1.5$, $\alpha = 10^{-4}$ models.  The width of the transition region widens, and the disc mass increases from left to right and from top to bottom, respectively.}
    \label{fig:tromb_h01_p15}
\end{figure*}

Figs. \ref{fig:tromb_h0025_p05} - \ref{fig:tromb_h01_p15} show the evolution of $\delta \Sigma$ in different $M_\mathrm{d} - \Delta r_\mathrm{dz}$ sets $\alpha=10^{-4}$.

\bsp	
\label{lastpage}
\end{document}